\documentclass{aa}  

\usepackage{graphicx}
\usepackage{txfonts}
\defcitealias{MartinezLombilla2019a}{ML19}

\usepackage{xcolor}
\definecolor{cincinnati-red}{RGB}{190,0,0}

\begin{document}

   \title{The truncation of the disk of NGC~4565:}

   \subtitle{Detected up to $z=4$~kpc, with star formation, and affected by the warp}


   \author{Cristina Mart{\'i}nez-Lombilla\inst{1,2,3,4}
          \and
          Ra{\'u}l Infante-Sainz\inst{1,2,5}
          \and
          Felipe Jim{\'e}nez-Ibarra\inst{6,3}
          \and
          Johan H. Knapen\inst{1,2}
          \and
          Ignacio Trujillo\inst{1,2}
          \and
          S{\'e}bastien Comer{\'o}n\inst{2,1}
          \and
          Alejandro S. Borlaff\inst{7,8,9}
          \and
          Javier Rom{\'a}n\inst{1,2,10}
          }

   \institute{Instituto de Astrof{\'i}sica de Canarias  (IAC), La Laguna, 38205, Spain
              \email{c.martinezlombilla@unsw.edu.au}
         \and
             Departamento de Astrof{\'i}sica, Universidad de La Laguna (ULL), E-38200, La Laguna, Spain
        \and
            School of Physics, University of New South Wales, Sydney, NSW 2052, Australia
        \and
            Australian Research Council Centre of Excellence for All-Sky Astrophysics in 3 Dimensions (ASTRO 3D), Stromlo, ACT 2611, Australia
        \and
            Centro de Estudios de Física del Cosmos de Aragón (CEFCA), Plaza San Juan 1, 44001, Teruel, Spain
       \and
          School of Physics $\&$ Astronomy, Monash University, Clayton, VIC 3800, Australia    
        \and
           NASA Ames Research Center, Moffett Field, CA 94035, USA
        \and
        Bay Area Environmental Research Institute, Moffett Field, California 94035, USA
        \and 
        Kavli Institute for Particle Astrophysics \& Cosmology (KIPAC), Stanford University, Stanford, CA 94305, USA
        \and
        Kapteyn Astronomical Institute, University of Groningen, Landleven 12, 9747 AD Groningen, The Netherlands
             }

   \date{Received XXX; accepted XX}

 
  \abstract
   {The hierarchical model of galaxy formation suggests that galaxies are continuously growing. However, our position inside the Milky Way prevents us from studying the disk edge. Truncations are low surface brightness features located in the disk outskirts of external galaxies. They indicate where the disk brightness abruptly drops and their location is thought to change dynamically. In previous analyses of Milky Way-like galaxies, truncations were detected up to 3~kpc above the mid-plane but whether they remain present beyond that height remains unclear.} 
   {Our goal is to determine whether truncations can be detected above 3~kpc height in the Milky Way-like galaxy NGC~4565, thus establishing the actual disk thickness. We also aim to study how the truncation relates to disk properties such as star formation activity or the warp.}
   {We perform a vertical study of the disk of NGC~4565 edge in unprecedented detail. We explore the truncation radius at different heights above/below the disk mid-plane ($0<z<8$~kpc) and at different wavelengths. We use new ultra-deep optical data ($\mu_{g,\rm{lim}}=30.5$~mag~arcsec$^{-2}$; $3 \sigma$ within $10 \times 10$~arcsec$^{2}$ boxes) in the $g$, $r$ and $i$ broad bands, along with near- and far-ultraviolet, H$\alpha$, and \ion{H}{i} observations.}
   {We detect the truncation up to 4~kpc in the $g$, $r$ and $i$ ultra-deep bands which is 1~kpc higher than in any previous study for any galaxy. The radial position of the truncation remains constant up to 3~kpc while higher up it is located at a smaller radius. This result is independent of the wavelength but is affected by the presence of the warp.}
   {We propose an inside-out growth scenario for the formation of the disk of NGC~4565. Our results point towards the truncation feature being linked to a star-forming threshold and to the onset of the disk warp.}

   \maketitle
%

\section{Introduction} \label{section:Intro}

Over several decades, the analysis of internal secular processes, star formation history, and the interstellar medium \citep[e.g.][]{Kormendy2004} has enabled significant advances in our understanding of the evolution of galaxies within their optical radius. In contrast, our knowledge of the outskirts of galaxies —the so-called low surface brightness (LSB) science— has come from a recent set of observational studies and theories that have provided precise details on the hierarchical nature of cosmological structure formation \citep[e.g.][]{Martinez-Delgado2009, Kim2012, Duc2015, Mihos2017, Kaviraj2017, KadoFong2018, Zaritsky2019, Iodice2019, Montes2019, Trujillo2021, Montes2022, Saifollahi2022}.

One aspect of this advance relates to galaxy disk truncations, which constitute a well-defined and abrupt drop in the surface brightness typically observed at radial distances of $\sim$4 times the exponential scale length of the inner disk \citep{vanderKruit1979, vanderKruit1981a, vanderKruit1981b}. These truncations are found in three quarters of the thin disks in spiral galaxies \citep{KruitFreeman2011, Comeron2012}. However, the nature of disk truncations remains unclear and different scenarios have been proposed to explain their origin. Some researchers explain the truncation as the location of those stars having the largest angular momentum at the moment of the protogalaxy collapse \citep[][]{vanderkruit1987a}, while others have suggested that truncations are associated with a threshold in star formation \citep{Kennicutt1989, Bakos2008, Rovskar2008b, Rovskar2008a, MartinezLombilla2019a, DiazGarcia2022}.

The location of the truncations at the edge of disks and therefore within the LSB regime (i.e. $\mu_{V} > 24$~mag~arcsec$^{-2}$) makes these features difficult to detect at large heights above the mid-plane. In consequence, the truncations have historically been considered a property of the galactic mid-plane. However \citet[][hereafter \citetalias{MartinezLombilla2019a}]{MartinezLombilla2019a} studied the truncations of two edge-on nearby galaxies, NGC~4565 and NGC~5907, in a wide wavelength range from the near ultraviolet (NUV) to the near-infrared (NIR) at different altitudes above/below the galaxies disk mid-plane. They found that the radial location of the truncation is independent of both wavelength and altitude above the galaxies' mid-plane. The truncation was detected as high as 3~kpc height above the galaxies mid-plane. In addition, \citetalias{MartinezLombilla2019a} associated the truncation location with a star formation threshold. At the location of the truncation, they obtained a face-on deprojected stellar mass density of 1--2~M$_{\odot}$~pc$^{-2}$, very close to the critical gas density necessary to transform gas into stars ($\sim 3-10~ M_{\odot}$~pc$^{-2}$), indicating an efficiency of $\sim$30 percent in transforming gas into stars. Independently, \cite{Gilhuly2020} confirmed the existence of such a star formation threshold in NGC~4565 although they proposed an accretion-based build-up of the outer disk. Recently, \cite{DiazGarcia2022} analysed the disk of UGC~7321, a well-studied edge-on, low-mass, diffuse, isolated, bulgeless, and ultra-thin galaxy. They reported the discovery of a truncation at and above the mid-plane ($\sim 0.5$~kpc height, i.e., where $90\%$ of the light comes from the thick disk) at all the probed wavelength ranges, from FUV to NIR. This truncation was also found to be linked to a star formation threshold. 

Based on the above results, \citet{Trujillo2020} proposed a physically motivated galaxy size indicator based on the location of the gas density for star formation threshold. As a proxy to identify this position, \citet{Trujillo2020} suggested using 1~M$_{\odot}$~pc$^{-2}$. This stellar mass density value refers to the region where the gas density threshold for star formation in galaxies is found theoretically \citep[see e.g.][]{Schaye2004}. \citet{Chamba2022} has explored at which stellar mass density value the truncation is found for a large number of galaxies. They found that 1~M$_{\odot}$~pc$^{-2}$ is a good approximation for galaxies changing several orders of magnitude in mass.

In this work, we present a follow-up study to \citetalias{MartinezLombilla2019a} in which we analyse the vertical structure of the disk of the edge-on Milky Way-like galaxy NGC~4565. Due to the limited depth of the data and image quality limitations, \citetalias{MartinezLombilla2019a} did not find the altitude above the galactic mid-plane at which the truncation disappears. Thus, an evaluation of the consequences that the vertical extent of the truncation could have on the structure and origin of the galaxy disk components is missing. In this study, we aim to address the issue using new ultra-deep optical data from the 4.2~m William Hershel Telescope (WHT, $\mu_{g,\rm{lim}}=30.5$~mag~arcsec$^{-2}$; $3 \sigma$ within $10 \times 10$~arcsec$^{2}$ boxes) in the $g$, $r$ and $i$ broad bands. We perform an unprecedented deep vertical analysis of the light and colour distribution of its disk. Our goal is to firmly establish the location of the truncation at any altitude and give a vertical extent of the thin and thick disks of NGC~4565. Based on those results we discuss the origin of the two disk components. In addition, we broaden the study with H$\alpha$ and \ion{H}{i} data, which allows us to analyse the stellar population and star formation rate (SFR) at the truncation and beyond. 

NGC~4565 is a barred edge-on spiral galaxy with a similar mass to that of the Milky Way ($88.5 \pm 0.5$~deg of inclination to the line-of-sight according to \citetalias{MartinezLombilla2019a}, and $v_{\rm{rot}} = 243.6 \pm 4.7$~km/s from \citealt{Makarov2014}). In terms of its mass assembly history, recent studies suggest that NGC~4565 has not experienced a strong interaction \citep{Gilhuly2020, Mosenkov2020} that disturbs the disk structure. \citet{Gilhuly2020} proposed a tidal ribbon event to explain the fan-like structure in the northwest part of the outer disk. The physical properties of NGC~4565 and the absence of evidence of a major disruption event guarantee a reliable comparison of the vertical properties of the disk of NGC~4565 to that of the Milky Way. In addition, the combination of its nearby location in the Coma I Group (\citealt{Gregory1977}; $\sim 13$~Mpc distance), and the size of its disk ($R \sim 30$~kpc; e.g., \citetalias{MartinezLombilla2019a}, \citealt{Gilhuly2020}), sets NGC~4565 as the galaxy of this kind with the most extended apparent size on the sky. This means that NGC~4565 provides the best spatial resolution ($\sim 65$~pc/arcsec) for a detailed study of a galaxy's vertical disk structure.

We have detected the highest truncation in a galaxy disk up to date. We measured the disk truncation up to 4~kpc above/below the NGC~4565 mid-plane, with a mean radial truncation position of $25.9~\pm~0.4$~kpc for the three $g$, $r$ and $i$ bands. This truncation feature is also found in the H$\alpha$ image at the galaxy mid-plane and in \ion{H}{i} data up to $\sim 1.5 - 2$~kpc height. Also, a colour analysis allowed us to measure the threshold of star formation associated with the truncation and the blue stellar populations of the disk warp. The results from the analysis of the truncation and the vertical disk structure of the edge-on Milky Way-like galaxy NGC~4565 are presented in this paper as follows. In Section~\ref{section:Data}, we describe the data in each wavelength range; we then explain the methods used to extract the information from that data in Section~\ref{section:Method}, giving details of key processes such as the mask construction, sky subtraction, PSF modelling, profile extraction, how we estimated the truncation radius and the star formation rates. The main results on the truncation radius and how they compare with previous works are presented in Section~\ref{section:Results}. We then discuss the implications of these results regarding the star formation activity at the edge of the disk in Section~\ref{section:SFRresults}, the actual vertical extent of the disk in Section~\ref{Sec:vertDisk}, and what is the role of the warp in the truncation location in this Milky Way-like galaxy in Section~\ref{sec:warpRole}. Finally, we draw our conclusions in Section~\ref{section:conclusions}. All magnitudes are provided in the AB system.


\section{Data} \label{section:Data}

\begin{figure*}
\centering
\includegraphics[width=\textwidth]{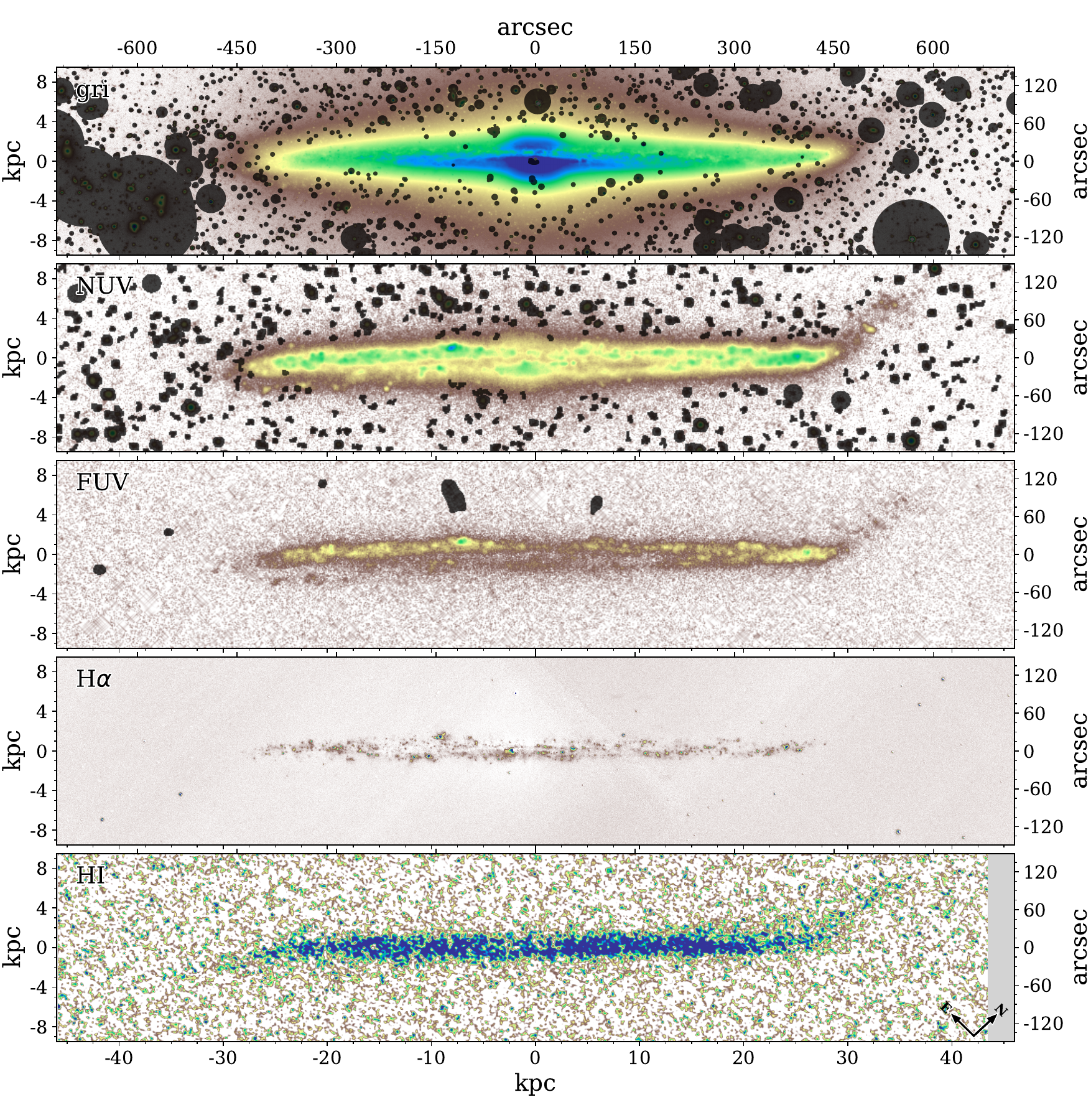}
  \caption{NGC~4565 images in the wavelength ranges studied in this work with our custom-made mask in shaded black. From top to bottom: coaddition of the optical $g$, $r$, and $i$ band images using our observed data at WHT; NUV image from GALEX; FUV image from GALEX; H$\alpha$ narrow-band image from observation at INT; and \ion{H}{i} integrated intensity map from VLA. The radial location of the disk edge of NGC~4565 is almost the same in all wavelength ranges.}
     \label{Fig:GalIm}
\end{figure*}

We present an ultra-deep photometric study of the vertical structure of the disk of the edge-on galaxy NGC~4565 (see Fig.~\ref{Fig:GalIm}). NGC~4565 is a well-known nearby galaxy whose structural component parameters make it a suitable analogue of our Milky Way \citep{BlandHawthorn2016, Kormendy2019}.

In this work, we used the same physical parameters for NGC~4565 as in \citetalias{MartinezLombilla2019a} (see their Table~1). The analysis was mainly performed in the optical wavelength range, including ultra-deep broad-band data in $g$, $r$, and $i$ filters, and an extra image in a narrow H$\alpha$ filter. We also used far- and near-ultraviolet (FUV and NUV respectively), as well as \ion{H}{i} data to extract additional and complementary properties. In the following sections, we explain how each type of data was obtained.

\begin{figure*}
\centering
\includegraphics[width=\textwidth]{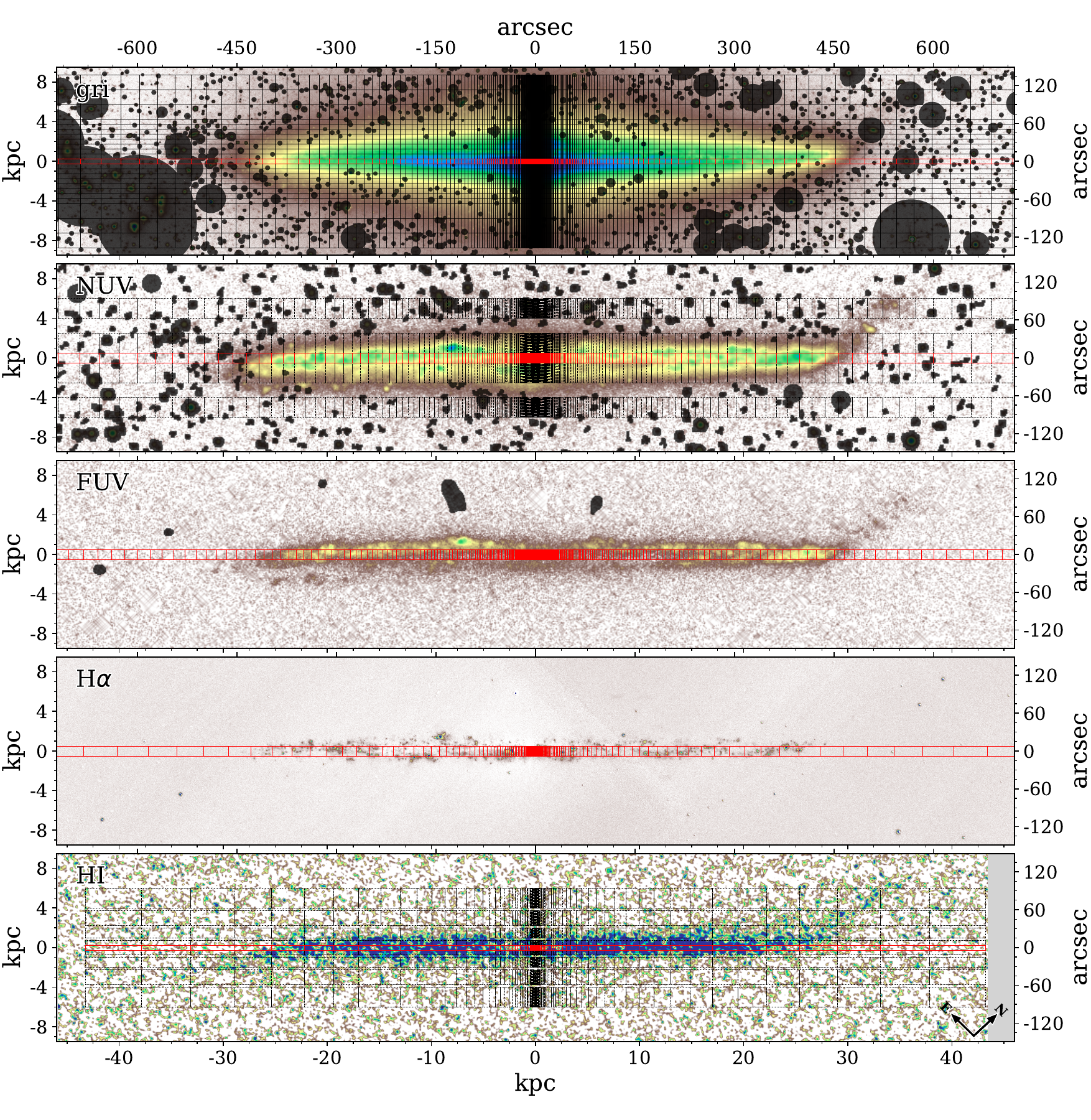}
  \caption{Same that Fig.~\ref{Fig:GalIm} but showing the bins of the extracted profiles. The bins extracted over the galaxy mid-plane are indicated in red while the ones above/below the disk mid-plane are shown in black. The readers checking the digital version of the paper can zoom in on the image to see even the smallest bin separately.}
     \label{Fig:GalIm_ap}
\end{figure*}

\subsection{New WHT ultra-deep broad-band optical data}  \label{subsection:WHT}

We obtained the ultra-deep imaging data at the 4.2~m William Herschel Telescope (WHT; La Palma, Spain). The images were obtained with the Physics of the Accelerating Universe Camera \citep[PAUCam;][]{Padilla2019} in three broad-band filters $g$, $r$, and $i$, with a total amount of time observed in each of them of 4.8, 3.1, and 4.4 hours, respectively. The large field of view of PAUCam ($\sim$1 $\times$ 1~deg$^{2}$) allows us to observe the target but also its surroundings making it possible to perform an exquisite data reduction and sky background treatment. The images have a pixel size of 0.253~arcsec, this is, a resolution of 16~pc~pix$^{-1}$ considering that the distance to the galaxy is 13.14~Mpc (distance obtained from the average of all calculated distances to NGC~4565 since 2010 in the NASA$/$IPAC Extragalactic Database\footnote{\url{http://ned.ipac.caltech.edu/}}, NED). We also observed very bright stars to construct an extended point spread function (PSF) and correct the scattered light field (see Sect.~\ref{subsection:NGC_PSFcorr}).

We designed an observational strategy and data reduction optimised to detect and preserve the low surface brightness structures around our target. When observing, we followed a dithering pattern with large steps (of the order of the size of the galaxy) combined with different rotation positions of the camera. This allows to correct systematic effects introduced by the instrument/telescope system and to perform a night sky flat-fielding correction procedure \citep[see e.g.,][]{TrujilloFliri2016, Montes2020}. 

We briefly summarise the processing of the raw data. We obtained individual images with an exposure time of 200~seconds each. The PAUCam consists of 14 CCDs, with 4 channels for each CCD. Due to the illumination effects on the more external CCDs, we decided to use only the 8 central CCDs with the best illumination quality. Each channel was bias-corrected and flat-fielded. The master flat field images were obtained from the science images themselves. The methodology was to mask all sources on each individual image using \texttt{NoiseChisel} \citep{Akhlaghi2015, Akhlaghi2019}, normalise those images, and combine them to obtain the master flat field image. To subtract the sky background we masked all the signal pixels detected by \texttt{NoiseChisel} and computed a very smooth surface model to the remaining background pixels. We then subtracted a 3-$\sigma$-clipped median value of the sky model from the science images. This procedure ensures the correction of systematics and the sky background while preserving the low surface brightness structures. Further details will be provided in Infante-Sainz et al. (in prep.).

After the correction of the systematic effects, we used the Astrometry\footnote{\url{http://Astrometry.net}} software \citep{Lang2010} to produce a first-order astrometric solution that was later improved with \texttt{SCAMP} \citep{Bertin2006}. Relative astrometry techniques are very difficult to implement in images with large dithering steps and, at the same time, obtaining an accurate astrometric solution is key in low surface brightness studies as small offsets in the alignment of the images will introduce a loss of signal-to-noise ratio (S/N) in the final stacked image.

The next step is the co-addition of all the aligned images to increase the S/N of the final very deep images of NGC~4565. We used a 3$\sigma$-clipped median to generate a final image per filter. Finally, the photometric calibration was done using SDSS DR12 \citep{Alam2015}.

The final images of NGC~4565 have a limiting surface brightness of 30.5, 29.9, and 29.3~mag~arcsec$^{-2}$ ($3 \sigma$; $10 \times 10$~arcsec$^{2}$) for the $g$, $r$, and $i$ filters, respectively. They were calculated using \texttt{NoiseChisel} \citep{Akhlaghi2015, Akhlaghi2019} and following the procedure in \citet{Borlaff2019} and in \citet{Roman2020}. This is, obtaining the standard deviation of the images where all the sources are masked (see Sect.~\ref{subsection:Mask}) and applying this method in each band. These surface brightness limits are, $\sim$3~mag~arcsec$^{-2}$ deeper than in \citetalias{MartinezLombilla2019a}. The zero point value to convert between instrumental magnitudes and calibrated magnitudes is fixed to 22.5~mag for the three bands.

\subsection{INT narrow-band H$\alpha$ data} \label{subsection:Halpha}

We took narrow-band images of the galaxy NGC~4565 using the Wide Field Camera (WFC), an optical mosaic camera located at the prime focus of the 2.5~m Isaac Newton Telescope (INT; La Palma, Spain). We used the narrow filter H$\alpha$ 6568/95 centred at 6567.98~${\rm{\AA}}$ with a FWHM 95.63~${\rm{\AA}}$ (and a transmission $>80\%$). We also got images in the $R$ broad-band filter to subtract the contribution of the continuum in the narrow-band images. The WFC has a pixel size of 0.33~arcsec and a field of view of 34~$\times$~34~arcmin$^{2}$, wide enough to cover our target diameter of 16.6~$\pm$~0.1~arcmin at the isophotal level 25 mag arcsec$^{-2}$ \citep[value for the $B$ band corrected for galactic extinction; HyperLeda,][]{Makarov2014}. 

We performed a basic data reduction of both the H$\alpha$ and $R$ images using \texttt{THELI}\footnote{\url{https://www.astro.uni-bonn.de/theli}} \citep{Erben2005, Schirmer2013}, a software to automatically reduce astronomical images. Once the images were completely reduced, we subtracted the sky background from all the images in both filters. To do that, we adjusted a 2D polynomial function to a version of the images where all the sources were completely masked.

Then, we aligned and combined all the H$\alpha$ images weighted by their exposure time into a final mosaic with the median value of each pixel obtained after applying a 3$\sigma$-clipping rejection algorithm. The total observing time on the source with the narrow-band filter was 16991.86~s. In the same way, we got the $R$ broad-band mosaic, with a total exposure time of 3010.88~s. 

Finally, we subtracted the continuum contribution to the H$\alpha$ flux. The $R$-band continuum images were scaled and subtracted from the H$\alpha$ filter image using a scaling factor of $R$/H$\alpha = 0.0675$. We refer the reader to \cite{Knapen2004} for further details about the method used to calculate this scale factor of the continuum image. There are strong light gradients in the observed data that could not be properly subtracted, particularly in the innermost $\sim 4$~kpc.

The continuum-subtracted image was photometrically calibrated using public available SDSS DR12 data \citep{Alam2015}, and found that the calibrated zero point of the continuum-subtracted H$\alpha$ mosaic is $5.9 \times 10^{-16}$~erg~s$^{-1}$~cm$^{-2}$. The final image is shown in the second to last panel in Fig.~\ref{Fig:GalIm}.

\subsection{VLA \ion{H}{i} data} \label{subsection:HI}

\ion{H}{i} integrated intensity maps of NGC~4565 were provided by \cite{Yim2014}. Yim's team obtained the raw data from the Very Large Array (VLA) archive, available at the National Radio Astronomy Observatory\footnote{\url{https://science.nrao.edu/facilities/vla/archive/index}} (NRAO). Then, they reduced them using the CASA (Common Astronomy Software Applications) software package \citep{McMullin2007} which was built for reprocessing the whole data set in a uniform procedure. We refer the reader to \citet[][]{Yim2014} for a more detailed description of the data acquisition and reduction process.

In Table~\ref{tabla:HI} we list the observing parameters and in the bottom panel of Fig.~\ref{Fig:GalIm} we show the \ion{H}{i} integrated intensity maps of NGC~4565 in units of Jy\,beam$^{-1}$\,km\,s$^{-1}$. The one-sided warp of NGC~4565, noted in previous studies \citep[e.g.,][]{Rupen1991, Oosterloo2007, Zschaechner2012, Radburn-Smith2014, Yim2014, Das2020, Gilhuly2020} and in our optical data (see Sect.~\ref{subsection:WHT}) is evident. We visually identified the \ion{H}{i} warp onset radius at $\sim 26-27$~kpc.

\begin{table}
\caption{NGC~4565 VLA \ion{H}{i} observing parameters. $\theta$ and $\Delta v$ are the angular and velocity resolutions, respectively.}             
\label{tabla:HI}      
\centering                          
\begin{tabular}{c c}        
\hline\hline                 
Parameter & NGC~4565 \\    
\hline                        
Array configuration &   BCD \\
$\theta$ [arcsec]$^2$   &   6.26 $\times$ 5.59 \\ 
$\Delta v$ [km s$^{-1}$]    &   20   \\ 
Total flux [Jy km s$^{-1}$] &   274   \\ 
Channel noise [K]   &   2.73  \\ 
Velocity range  [km s$^{-1}$]   &   970-1490  \\ 
\hline                              
\end{tabular}
\end{table}

\subsection{GALEX ultraviolet data} \label{subsection:UV}

We use the deepest available data from the \textit{Galaxy Evolution Explorer} \citep[\textit{GALEX},][]{Martin2005, Morrissey2007}. \textit{GALEX} images have a circular field of view (FOV) of 1.2~degrees, a pixel size of 1.5~arcsec, and a spatial resolution (FWHM) of 4.2~arcsec and 5.3~arcsec in the FUV and NUV channels, respectively. The effective wavelengths are 1516~${\rm{\AA}}$ (FUV) and 2267$~{\rm{\AA}}$ (NUV). The FUV image has an exposure time of 1693.05 seconds and \textit{GALEX} zero point magnitude $m_{0, \rm{FUV}} = 18.82$~mag. In the case of the NUV data we use the same image as in \citetalias{MartinezLombilla2019a} with a long exposure time of 12050.15 seconds, which allows for a surface brightness limit of $\mu _{AB}\sim30.5$~mag~arcsec$^{-2}$ (1$\sigma$), and $m_{0, \rm{NUV}} = 20.08$~mag. The second and third panels in Fig.~\ref{Fig:GalIm} show the NUV and FUV images, respectively.

\section{Methods} \label{section:Method}

As mentioned in the Introduction (Sect.~\ref{section:Intro}), this work is a follow-up study of \citetalias{MartinezLombilla2019a}. Here we used deeper images of NGC~4565 data and a wider wavelength range to answer the open questions from our previous work. These are: 1) what is the altitude above the galaxy mid-plane where the truncation disappears?; and 2) what is the amount of stars that are being formed at the very edge of the disk of NGC~4565 --this is, the specific star formation rate (SFR)?

Thus, most of the techniques applied in this study were previously developed as a semi-automatic code optimised for LSB data as described in \citetalias{MartinezLombilla2019a} (see their Sect.~3). The following sections will only detail the changes with respect to the previous work.

\subsection{Mask} \label{subsection:Mask}

The aim of the mask is to cover the flux from all the sources other than our target. However, this is a complicated task when working with data that reach LSB levels such as ours, where we detect structures at $\mu _{g}\gtrsim30$~mag~arcsec$^{-2}$. At these depths, even the light from very faint objects in the surroundings of NGC~4565 could contribute to the measured galaxy flux. Thus, a dedicated mask is key to obtaining reliable surface brightness measurements.

There are two main differences relative to the masking procedure of \citetalias{MartinezLombilla2019a}. The first is that here we used Python-based routines only as they are built from modular and object-oriented scripts, that allow for easy implementation of modifications and improvements. In particular, we used the tools provided by \texttt{Photutils} \citep{Bradley2020} to detect astronomical sources using image segmentation. The second difference is the two-step approach using a ``hot+cold'' combined mask \citep{Rix2004} as in e.g., \cite{Montes2018}, \cite{Montes2020}, or \cite{MartinezLombilla2023}. We first obtain a ``cold'' mask optimised to detect bright and extended sources. Then, over an image --masked with the ``cold'' mask-- we get a ``hot'' mask that accounts for faint and small objects. In order to smooth the noise and maximise the sensitivity of the algorithm, the images were filtered with 2D circular Gaussian kernels (FWHM sizes of $\sim$1.3~arcsec and $\sim$0.8~arcsec for the ``cold'' and ``hot'' masks respectively), prior to thresholding. The detection threshold is 1.1$\sigma$ above the background.

Once the ``hot+cold'' combined masks were finished, the visual check of the extent of the masks is done accordingly to \citetalias{MartinezLombilla2019a} (see their Sect.~3.1). The aim of this last step is to make sure the masks are covering all undesirable flux. In the case of the NUV data, we directly used the mask obtained in \citetalias{MartinezLombilla2019a}.

\subsection{Sky background treatment} \label{subsection:SkyBackgTreat}

\begin{figure}
\centering
\includegraphics[page=1,width=\hsize]{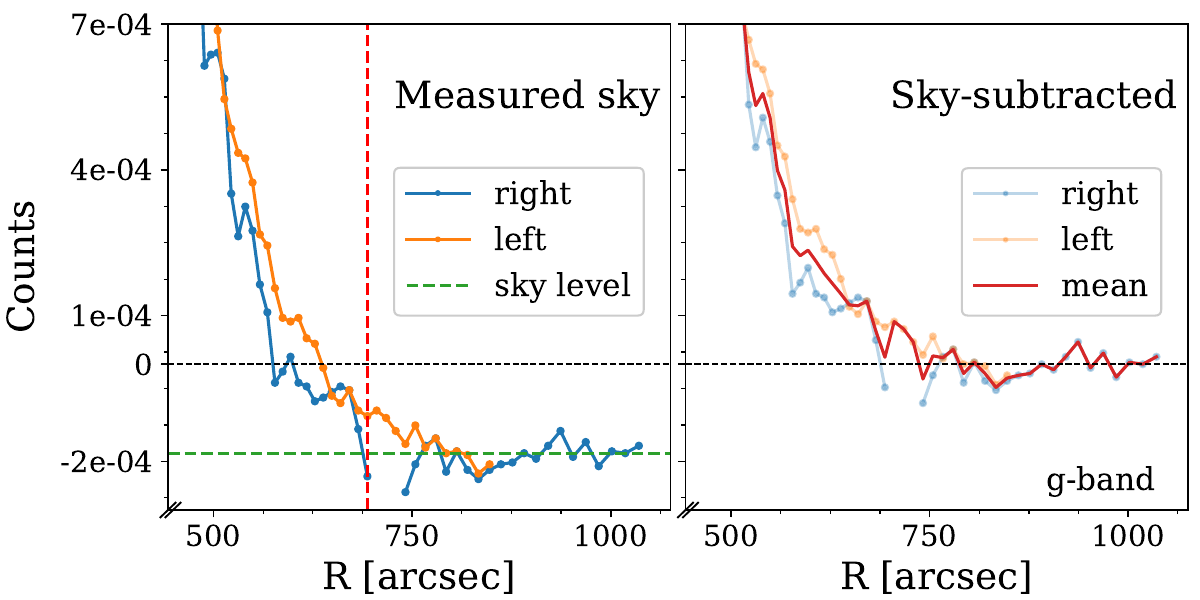}
  \caption{Local sky correction for the $g$ broad-band image. We show the outer parts of the profiles in detail ($R>450$~arcsec) along both sides of the galaxy disk. The profile extracted from the right side of the disk of NGC~4565 is in blue, and the one from the left side is in orange. \textit{Left panel}: Measured sky profiles. The vertical red dashed line indicates the radial distance from which we measured the mean sky value, the horizontal green dashed line is located at the measured mean sky level and the black dotted line shows the zero counts level. \textit{Right panel}: Sky-subtracted profiles, this is, the result of subtracting the mean sky value from the profiles on the left panel. The mean profile of the whole disk is in red solid line. The profiles are slightly corrected for over- and under-subtracted areas around our target.}
     \label{Fig:SkyCorr}
\end{figure}

We used co-added sky-rectified versions of all our images in which in general, a non-aggressive sky subtraction strategy had been done (see Sect.~\ref{section:Data}). In addition, we performed a second-order (also non-aggressive) local background subtraction. This is, we evaluated the sky covering only our region of interest in the outermost parts of the disk of NGC~4565. Following the procedures in \cite{PohlenTrujillo2006} and in \citetalias{MartinezLombilla2019a} (see details in their Sect.~3.2), we measured the mean background levels well beyond the disk edges through very extended radial surface brightness profiles. We consider as the sky the flattest part of the outermost region of the extended profiles ($R \gtrsim 700$~arcsec). We measured the mean values in counts of that sky region using a $3\sigma$-clipping rejection algorithm. We then subtracted the mean sky from the corresponding image. This local second-order background evaluation allows us to identify clearly where the surface brightness profiles of the disk of NGC~4565 reach the sky limit and in consequence, the location where the sky background starts to be dominant in the profiles. An example of the extracted profiles for our $g$ broad-band optical data is shown in Fig.~\ref{Fig:SkyCorr}.

\subsection{PSF correction in WHT broad-band optical data} \label{subsection:NGC_PSFcorr}

 The PSF characterises how the light coming from a point source is affected by the combination of the telescope, the instrument, the detector, and the atmosphere, and measures the extent of the scattered light \citep[e.g.][]{Michard2002, Slater2009, Sandin2015, TrujilloFliri2016, InfanteSainz2020}. Moreover, the outer wings of the PSF of bright stars can add some flux to the outer parts of extended objects or in low surface brightness objects around them. Thus, properties of thick disks, stellar haloes, and faint outskirts of galactic disks can be severely influenced by the wings of the PSF \citep[e.g.][]{Zibetti2004a, deJong2008, Sandin2014, Sandin2015, TrujilloFliri2016, Peters2017, Comeron2017, MartinezLombilla2019b}. For these reasons, is crucial to model not only the PSF of relatively faint and well-resolved stars but to construct our own very extended PSF that reaches the whole extent of those wings.

 We built three extended (R~$\sim$~20~arcmin) PSF models for each of the $g$, $r$, and $i$ filters of the PAUCam at the WHT telescope. Following the methodology outlined in \cite{Roman2020} and \cite{InfanteSainz2020}, we combined three very bright stars (saturated) with magnitudes 2.57, 4.82 and 5.84\,mag \citep[$G$-band from Gaia~EDR3,][]{Collaboration2021}, and 50 fainter stars of 11 and 13~mag. The resulting PSFs have full-width-at-half-maximum (FWHM) values of 0.8~arcsec in each of the $g$, $r$, and $i$ bands. We fit and subtracted all stars in the field-of-view of the camera brighter than $G=16$~mag, according to the Gaia~EDR3 catalogue, in order to minimise contamination in our photometry. To do so, we followed the steps in Sect.~3.2 of \cite{InfanteSainz2020}. These scattered light-subtracted images were the ones used in the following analysis in this work. 

 The extraordinary depth of the images as well as the extended PSFs characterisation allowed us to perform a detailed and reliable study of the outermost parts of the disk of NGC~4565, including colour and stellar population analyses above/below the mid-plane.

 As explained above, the physical properties derived from the outskirts of a galactic disk can be strongly affected by scattered light. Besides this, that PSF effect has been proven to be more dramatic in edge-on than in face-on galaxies, also affecting their outer colours \citep[e.g.,][]{deJong2008, Sandin2014, Sandin2015}. The strength of the PSF effect is also highly correlated with the depth of the images \citep[e.g.,][]{TrujilloFliri2016, Comeron2017, MartinezLombilla2019b}. In consequence, it is not only necessary to correct the scattered light field due to the bright stars, but also to the scattered light from the galaxy itself. Thus, NGC~4565 has also to be modelled and corrected.

 Interestingly, \citetalias{MartinezLombilla2019a} found that the effect of the PSF only slightly influenced the radial flux distribution and the slope of the surface brightness profiles of the edge-on galaxies NGC~4565 and NGC~5907. Also, in the same work they verified that the radial position of the galaxy truncations was not affected by the PSF (see their Sect.~4.3 and Figures~1 and 4). \cite{Borlaff2017} reported a similar result in a study of the disk breaks in a sample of Type-III S0 and E/S0 galaxies at $0.2 < z < 0.6$. They found that the PSF tends to increase the scale lengths of the inner and outer disk profiles, but it does not significantly affect either the central surface brightness values of the inner and outer disks or the break location. However, \citet{MartinezLombilla2019b} showed that a careful PSF treatment is absolutely indispensable for deep imaging of extended objects because if the PSF effect is not accounted for, flux and mass measurements of the outskirts of disks of edge-on galaxies can be overestimated when reaching surface brightness values of $\sim$28~mag~arcsec$^{-2}$ or deeper. In particular, they found that the mass of a thick disk can be overestimated by a factor of 1.5--2 in low-mass sources ($\sim 10^9 M_\sun$) and of 2.5--4 in intermediate- to high-mass galaxies ($> 10^{10} M_\sun$). Previous works such as \cite{Comeron2017} did not find this mass excess because they only reached surface brightness levels of $\sim$26~mag~arcsec$^{-2}$. Although the sample of five galaxies of \cite{MartinezLombilla2019b} is small, it is worth checking whether the PSF is now affecting the location of the truncation in our ultra-deep WHT optical images, especially in the highest (and faintest) parts of the disk of NGC~4565.

 We had the scattered light-subtracted images of WHT optical data in the $g$, $r$, and $i$ broad bands in which the light of the foreground stars had been corrected. However, it is still necessary to model the effect of the PSF over the galaxy itself. To do that, we use \textsc{imfit}\footnote{Precompiled binaries, documentation, and full source code (released under the GNU Public License) are available at the following website: \url{https://www.mpe.mpg.de/~erwin/code/imfit/} } \citep{Erwin2015}, which allows us to model the intrinsic light distribution of NGC~4565 and convolve it with the image of the extended PSF. Then, the PSF-convolved model of the galaxy is fitted to the observed data. In this way, a new image of the source is built as the result of the PSF-deconvolved model of NGC~4565 plus the PSF-convolved fitting residuals. These residuals consider all the non-symmetric features, such as the spiral arms, that the model cannot properly fit. A comprehensive overview of the steps, assumptions, and optimization algorithms used in this 2D deconvolution process have already been addressed in previous works \citep[e.g.][]{TrujilloFliri2016, Peters2017, MartinezLombilla2019b}.
 
 To model NGC~4565 we had to use four galaxy components, as indicated by its morphology \citep[][]{Buta2015}: a bulge using an elliptical 2D S{\'e}rsic function with generalised ellipses (“boxy” to ``disky'' shapes) for the isophotes instead of pure ellipses \citep{Athanassoula1990}; a bar represented by a 2D S{\'e}rsic function \citep{Sersic1968}; a disk truncated at the same distance as the data represented with a 3D broken exponential disk function; and finally, a halo component with a purely exponential function \citep[for more details about the analytical functions, see][]{Erwin2015}.

 An example of the final 2D deconvolved model of NGC~4565 and its components is shown in Fig.~\ref{Fig:model-profile} (Appendix~\ref{appendix:PSFmodels}). We see that the PSF is clearly affecting the outermost regions of the disk of NGC~4565 as the width of the galaxy changes after the deconvolution process. This is an effect clearly visible due to the extraordinary depth of our images. By using the scattered light-corrected images and these 2D PSF-deconvolved models of the galaxy, we are in the position to obtain reliable photometry which impacts the subsequent analysis of the physical properties of the outskirts regions in NGC~4565. We, therefore, use the images of the 2D PSF-deconvolved models of the galaxy for further analysis in this work. Hereafter, we refer to the 2D PSF-deconvolved model of the galaxy in each of the bands as ``the galaxy model''.

\subsection{Profile extraction} \label{subsection:SBPExtraction}

\subsubsection{Broad-band optical profiles} \label{subsubsec:Opticalprof}

Our goal is to measure the height above the mid-plane where the disk truncation disappears. For that, we need to extract radial surface brightness profiles (RSBP) at different altitudes with the best possible vertical resolution and up to the highest regions that the S/N of our images allows. A galaxy RSBP can be extracted by calculating the fluxes through a slit with a given width along the radial axis. In general, we followed the method explained in \citetalias{MartinezLombilla2019a} (see details in their Sect.~3.3) over the galaxy models (details in Sect.~\ref{subsection:NGC_PSFcorr}). However, the depth of our new data allowed us to implement some statistical improvements as well as a slightly different slit configuration. These changes are described below:

\begin{itemize}
    \item For each bin, we calculated the surface brightness by applying the Galactic absorption coefficient correction at each wavelength from the NED \citep{Schlafly2011}. For simplicity, we applied the same correction coefficient throughout the whole galaxy ($A_g =0.051$, $A_r =0.035$, and $A_i =0.026$~mag). Each resulting surface brightness value was obtained by calculating the 3$\sigma$-clipping median of the given set of pixels.
    
    \item We increased the number of bins to 180 of each RSBP due to higher S/N of the images (in \citetalias{MartinezLombilla2019a} there were 150 bins in each RSBP). The radial spacing between bins remains  evenly spaced on a radial logarithmic scale.

    \item The uncertainties for each bin were defined as in \citetalias{MartinezLombilla2019a} but this time the $RMS$ quantity was determined over 25000 well-distributed and different background regions of $\sim 10 \times 10$~arcsec$^2$ in the masked image.
    
    \item The new ultra-deep optical data from the WHT allowed us to reach higher altitudes above and below the NGC~4565 mid-plane. Thus, the widths and vertical locations of the \textit{shifted} RSBP are slightly different from the \citetalias{MartinezLombilla2019a} configuration. Now, the boxes with 0.5~kpc height reach up to 4~kpc. Then we have two steps of 1.0~kpc at 5.0 and 6.0~kpc in height, and one last step of 1.5~kpc at an 8.0~kpc altitude. In this way, we reach the same external regions as in \citetalias{MartinezLombilla2019a} but with better spatial resolution and higher S/N allowed by our data. 
    
    \item Our technique provides reliable RSBPs down to $\mu _{g}=$~30.5~mag~arcsec$^{-2}$ ($3 \sigma$; $10 \times 10$~arcsec$^{2}$).
    \end{itemize}
    
 The size and location of all the apertures and their corresponding bins for the broad-band optical profiles are shown in the first panel of Fig.~\ref{Fig:GalIm_ap}.

\subsubsection{H$\alpha$ profiles} \label{subsubsec:Halphaprof}

The H$\alpha$ stacked and continuum-subtracted image needs a different treatment in terms of galactic extinction correction. For this narrow filter, it is also required to remove possible contributions to the total flux from other emission lines. In order to extract RSBPs in H$\alpha$ we undertook the following steps:

\begin{itemize}
    \item  To correct for total absorption for the galaxy, we needed to determine the contribution from both, the foreground Galactic absorption in the $R$ band, $A(R)$, and the internal absorption of NGC~4565 in the H$\alpha$ narrow filter, $A(\rm{H}\alpha)$. Thus, the total absorption in magnitude units can be obtained as the sum of them:  $A_{T} = A(R) + A(\rm{H}\alpha)$. For $A(R)$, we used the value given by the NED \citep[from][]{Schlafly2011}. In the case of $A(\rm{H}\alpha)$, several solutions have been proposed in the literature \citep[e.g.,][]{Kennicutt1983, Niklas1997, Helmboldt2004, James2005}. In this work we took the approach of \citet[see their Sect.~5.7]{Sanchez-Gallego2012} where $A(\rm{H}\alpha)$~=~0.3 when $M_{B}~\leq -16$ \citep[the absolute \textit{B}-band magnitude of NGC~4565 is -21.44~$\pm$~0.24; HyperLeda,][]{Makarov2014}.
        
    \item As our narrow-band filter is wider than 35-40~${\rm{\AA}}$, we also applied the statistical correction to the H$\alpha$ fluxes for [\ion{N}{ii}] forbidden lines contamination at $\lambda$6548, 6584~${\rm{\AA}}$. We estimated this contribution using the expression from \cite{Kennicutt2008}: $\rm{log([\ion{N}{ii}]/H\alpha)}$~=~0.54 if $M_{B} \leq $-21. As the FWHM of the H$\alpha$ 6568/95 filter is 95.63~${\rm{\AA}}$ there is no need to correct the [\ion{N}{ii}]/H$\alpha$ ratio for the transmission profile of the filter \citep{Sanchez-Gallego2012}.
        
    \item After all the above corrections, we extracted one RSBP along the galaxy mid-plane in a region of 1~kpc width in physical flux units (see the second to last panel in Fig.~\ref{Fig:GalIm_ap}). This was the best spatial resolution and the higher height we could reach with the current data. However, the measured flux in the inner galaxy disk ($\lesssim 4$~kpc) is slightly overestimated as we could not properly subtract the sky emission and the continuum due to strong light gradients in the observed data.
\end{itemize}

\subsubsection{\ion{H}{i} profiles} \label{subsubsec:HIprof}

We extracted RSBPs from the \ion{H}{i} data as shown in the bottom panel of Fig.~\ref{Fig:GalIm_ap}. One profile was extracted along the galaxy mid-plane and the rest at four altitudes above/below it: at 0.5, 1.5, 3, and 5~kpc. The profiles have an increasing width from 0.5~kpc to 2~kpc the higher the position above/below the galaxy mid-plane. However, due to the nature of these data, the procedure is slightly different from the one explained above for the data in other wavelength ranges.

Our VLA zeroth-moment intensity maps are in Jy\,beam$^{-1}$\,km\,s$^{-1}$ units but we need to extract a surface mass density profile, this is, in units of M$_{\odot}$\,pc$^{-2}$. To do that, we first converted the value of each bin of the RSBP from flux density per area in Jy\,beam$^{-1}$ to brightness temperature in K, by specifying the beam area. We used the Rayleigh-Jeans equivalent temperature as it shows a linear relation between flux and temperature. This equivalence is usually known as “Antenna Gain” as the flux density sensitivity at a given frequency is related to the aperture size, while the telescope brightness sensitivity is not. Thus, the Rayleigh-Jeans relation is only dependent on the aperture size. For a more comprehensive explanation, see equations 8.16 and 8.19 in \citet{Wilson2009}.

The derived values in K\,km\,s$^{-1}$ were converted to \ion{H}{i} surface mass densities using the optically thin approximation from \cite{Hunter2012}:    

\begin{equation}
N{\rm (HI) \,\, [atoms\,cm^{-2}]} = 1.823 \times 10^{18} I{\rm _{HI} \,\, [K~km~s^{-1}]} \,.
\end{equation}

This, together with the mass of the \ion{H}{i} atom value and the corresponding unit adjustments, returns the surface mass density values in M$_{\odot}$~pc$^{-2}$. The uncertainties for each bin of the profiles reflect the standard deviation of the fluxes within each bin. These surface mass density measurements are not corrected for the inclination of the galaxy. The \ion{H}{i} profiles are shown in Fig.~\ref{Fig:Hi-panel}.

\subsubsection{UV profiles} \label{subsubsec:NUVprof}

The NUV data has poorer spatial resolution and S/N than optical. For this band we enlarged the height of the uppermost bins to increase the S/N (see the second panel in Fig.~\ref{Fig:GalIm_ap}). We extracted only 3 profiles, at the mid-plane (of 1~kpc width), at 1.5, and 5~kpc above and below it (the latter two of 2~kpc width). There are two main reasons why we adopted this criterion: first, the NUV data has already been studied in \citepalias[][]{MartinezLombilla2019a}; and second, because we only use the NUV image to compare the profiles at three particular vertical locations in Sect.~\ref{subsection:TrucPosresults}. 

The FUV data is used for qualitative comparison purposes on the disk light emission in different regions. Therefore, we only extracted one RSBP along the galaxy mid-plane in a region of 1~kpc width.

\subsection{Truncation position} \label{subsection:TruncationPosition}

Once we extracted all the RSBPs at all the different highs above and below the galaxy mid-plane and in all the wavelengths, the last step of the process was to determine the position of the truncation in the radial axis. We define the truncation as the sharp edge in the disk of a highly-inclined galaxy \citep[although some works have been able to detect truncations in face-on galaxies; e.g.,][]{Martin-Navarro2012, Peters2017, Trujillo2021}. This sharp edge is seen in a RSBP as the change in the slope, mimicking a type II break but in the galaxy disk edge. In consequence, the regions before and after the truncation can be modelled with straight lines of different slopes.

We developed a routine to detect the truncation. This routine fits two exponential functions to the $g$, $r$, $i$ and \ion{H}{i} RSBPs to fit the regions of the disk  inside and outside the truncation. When working in logarithmic scales of flux units, these functions are seen as lines that intersect at a given point. That point is considered as the truncation radial location. The fitting of exponential functions is restricted to either the inner or the outer part of the disk. Those boundaries are estimated by a visual inspection using an interactive interface of the routine. However, with the aim of reaching unbiased results, we carried out a fitting procedure afterwards based on Sequential Least Squares Programming (SLSQP) optimisation algorithm and least squares statistic. These algorithms perform a broken linear fit of the two lines and find the intersection point if any. This fitting procedure is combined with an iterative 3$\sigma$-clipping outlier removal technique in which, given a maximum number of iterations (3 in our case), outliers are removed and the fitting is performed for each iteration until no new outliers are found or the maximum number of iterations is reached. In addition, the routine performs a Bootstrap re-sampling over 100 fitting points subsets of each of the two lines in each RSBP. Bootstrap re-sampling is a method that estimates the variability of our results by making a more detailed analysis of the parameter distributions. The combined set of bootstrapped parameter values can be used to estimate confidence intervals, and consequently, to discern between the multiple good solutions.

The derived truncation radii, obtained as the location where both lines intersect, are almost always in agreement with those derived by eye. We computed the errors as the standard deviation of the determination of the truncation position distribution for a given band and height.

\subsection{Star formation rate (SFR)} \label{subsection:SFR}

H$\alpha$ is one of the preferred tracers of star formation as it is a strong emission line with a short timescale ($\lesssim 10$~Myr, \citealt{Hao2011}), relatively easy to observe, and with a good spatial resolution. However, this narrow wavelength range also presents many uncertainties and difficulties due to extinction, [\ion{N}{ii}] contamination, diffuse/absorbed fractions, or sensitivity to the upper initial mass function (IMF) slope in the very upper mass range. So, in order for H$\alpha$ to be a precise tracer of the SFR one must take into account extinction. Moreover, when using narrow-band filters around H$\alpha$ to measure the luminosity of the line, it is required to consider the contamination of the H$\alpha$ flux by the neighbouring [\ion{N}{ii}] $\lambda$~6548 and 6584 lines. These two corrections were addressed in this work (see Sect.~\ref{subsubsec:Halphaprof}).

The SFR can be obtained from the measured H$\alpha$ fluxes. We applied the following relation between the H$\alpha$ line intensity and the SFR from \cite{Kennicutt2009}:  

\begin{equation}
{\rm SFR[M_{\odot} \, yr^{-1}]} = 5.5 \times 10^{-42} { L(\rm{H}\alpha)} \,\, , 
\end{equation}

\noindent
with $L$(H$\alpha$) being the luminosity, determined as follows:

\begin{equation}
{L(\rm{H}\alpha)[erg \, s^{-1}]} = 4\pi D (3.086 \times 10^{24})^{2} { F_{\rm{H}\alpha}} \,\, ,
\end{equation}

\noindent
where $D$ is the distance to the galaxy in Mpc and $F_{\rm H\alpha}$ is the flux in H$\alpha$ corrected by absorption and [\ion{N}{ii}] contamination. For the estimation of this SFR conversion factor, we assumed a ``Kroupa'' IMF \citep{Kroupa2003}. 

\section{Disk truncation radius results and comparison with previous works} \label{section:Results}

In the following sections, we highlight the main outcomes derived after applying the methods described in Sect.~\ref{section:Method} and how these findings fit within the current context of the field. We have used the deepest available data of the edge-on Milky Way-like galaxy NGC~4565 in a broad wavelength range, to measure the disk truncation radius at different heights above/below the mid-plane. To do that, we have extracted multiple radial surface brightness profiles of the disk of NGC~4565 as shown in Fig.~\ref{Fig:GalIm_ap}. Then, we modelled those profiles and fitted the truncation location (see Sect.~\ref{subsection:TruncationPosition}). In the following sections, we detail the results derived in this work and compare them with previous relevant studies.

\begin{figure}
\centering
\includegraphics[width=\hsize]{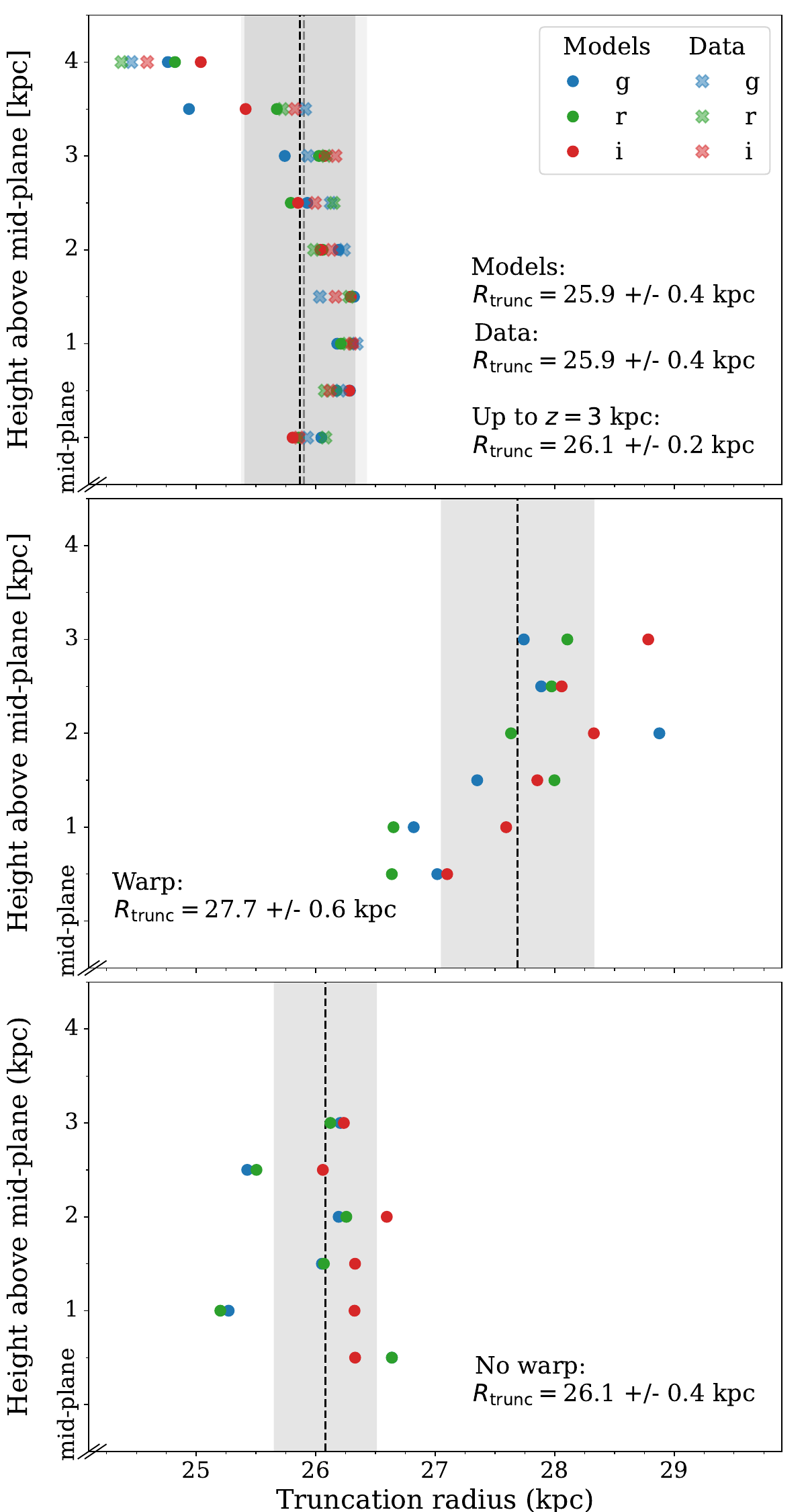}
  \caption{Truncation radius measured from the RSBP along the disk mid-plane (i.e. 0~kpc), and at different heights above/below the NGC~4565 mid-plane (the horizontal axis is scaled so as to show the region around the truncation). The shaded regions denote the standard deviations of the corresponding distribution of truncation positions. The vertical dashed lines represent the mean values of the corresponding truncation radius. \textbf{\textit{Top:}} broad-band optical data values in comparison with the corresponding models. \textbf{\textit{Centre:}} truncation radii for the disk quadrant with warp (north quadrant or upper right in Fig.~\ref{Fig:quadrants}) for the three $g$, $r$, and $i$ bands. \textbf{\textit{Bottom:}} same as in the \textit{centre} panel but for a disk quadrant with no warp feature (west quadrant or lower right in Fig.~\ref{Fig:quadrants}). In this work we detected truncations up to 4~kpc above/below the galaxy mid-plane, that is, 1~kpc higher than in \citetalias{MartinezLombilla2019a}. The presence of a warp clearly affects the radial location of the truncation feature.  }
     \label{Fig:TruncPosBoth}
\end{figure}

\subsection{Truncation detected up to 4~kpc above/below the NGC~4565 mid-plane} \label{subsection:TrucPosresults}

We obtained a mean radial truncation position of 25.9~$\pm$~0.4~kpc for the three $g$, $r$, and $i$ bands and for all the altitudes above/below the galaxy mid-plane, up to 4~kpc. Beyond this height, the truncation is not detected. These findings are shown in the top panel of Fig.~\ref{Fig:TruncPosBoth}, where we put together the values of the radial truncation positions obtained from the RSBPs of the galaxy models and those of the optical broad-bands (see Sect.~\ref{subsection:TruncationPosition}). The full set of RSBPs above/below the galaxy mid-plane are shown in the Appendix~\ref{appendix:RSBP} while in Table~\ref{tabla:scalelenghts} are the values of the corresponding scale lengths before and after the truncation obtained in the fittings of the profile. In Figs.~\ref{Fig:g-panel}, \ref{Fig:r-panel}, \ref{Fig:i-panel}, and \ref{Fig:Hi-panel} we have the RSBPs extracted from the observed data, while Figs.~\ref{Fig:g-model-panel}, \ref{Fig:r-model-panel} and \ref{Fig:i-model-panel} show the ones extracted from the models of NGC~4565. As mentioned in Sect.~\ref{subsection:NGC_PSFcorr}, we are going to use the optical data from the latter for further results and discussion. The error bars in the profiles account for the uncertainties in the source detection, the detector noise, and the sky fluctuations.

\begin{figure*}
\includegraphics[width=\textwidth]{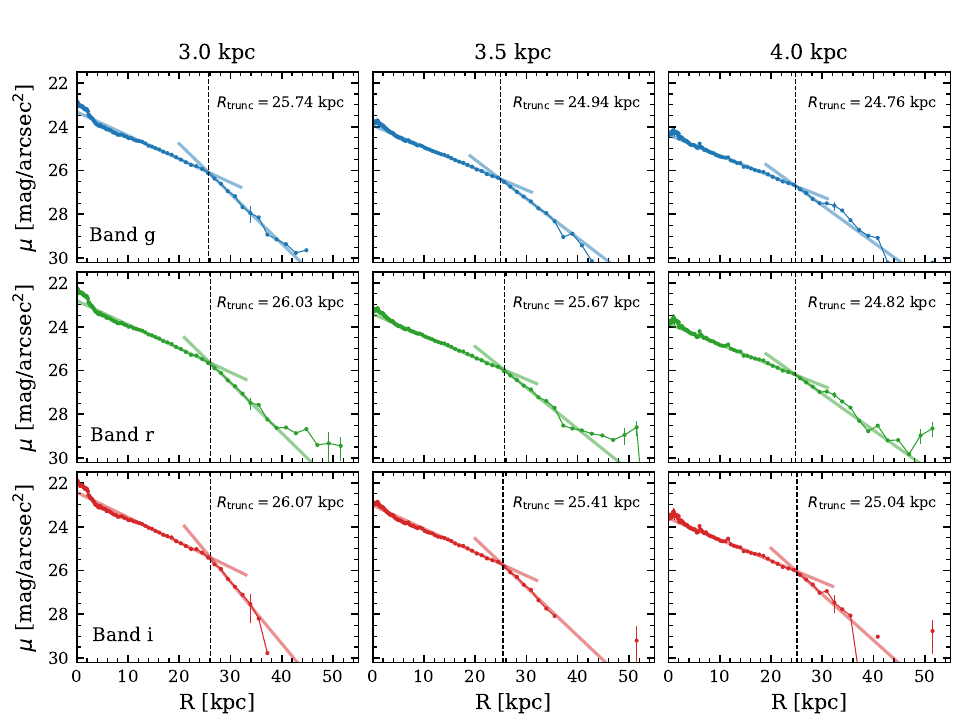}
\caption{Detail of the two linear fits (transparent straight lines) before and after the truncation radius. These lines fit the radial surface brightness profiles extracted at 3, 3.5 and 4~kpc height above/below the galaxy mid-plane (\textit{columns}) for each ultra-deep $g$ (blue), $r$ (green), and $i$ (red) band (\textit{rows}). The surface brightness profiles are those from the PSF-deconvolved models of NGC~4565 shown in Figs.~\ref{Fig:g-model-panel}, \ref{Fig:r-model-panel}, and \ref{Fig:i-model-panel}. In each panel, the intersection radius between the two lines provide the truncation radius of the corresponding profile, which is indicated with a dashed vertical line. The truncation radius value is indicated in the top right corner.}
\label{Fig:Fit_z3}
\end{figure*}

The lower S/N of the data in \citetalias{MartinezLombilla2019a}, limited their measurements of any reliable radial position of the truncation of the disk above 3~kpc in height. However, in this work we are confidently detecting the truncation up to 4~kpc in the ultra-deep optical data, this is, at 1~kpc higher locations than in any previous study for any studied galaxy. These high-altitude off-plane truncation detections are shown in detail in Fig~\ref{Fig:Fit_z3}. Beyond 4~kpc the truncation is not detected while the uncertainties of the surface brightness measurements are smaller than the fluctuations of the profiles. This guarantees that the data provides enough S/N at high altitudes above the galaxy mid-plane, ensuring the non-detection of the truncation beyond 4~kpc.

\begin{figure*}
\includegraphics[width=\hsize]{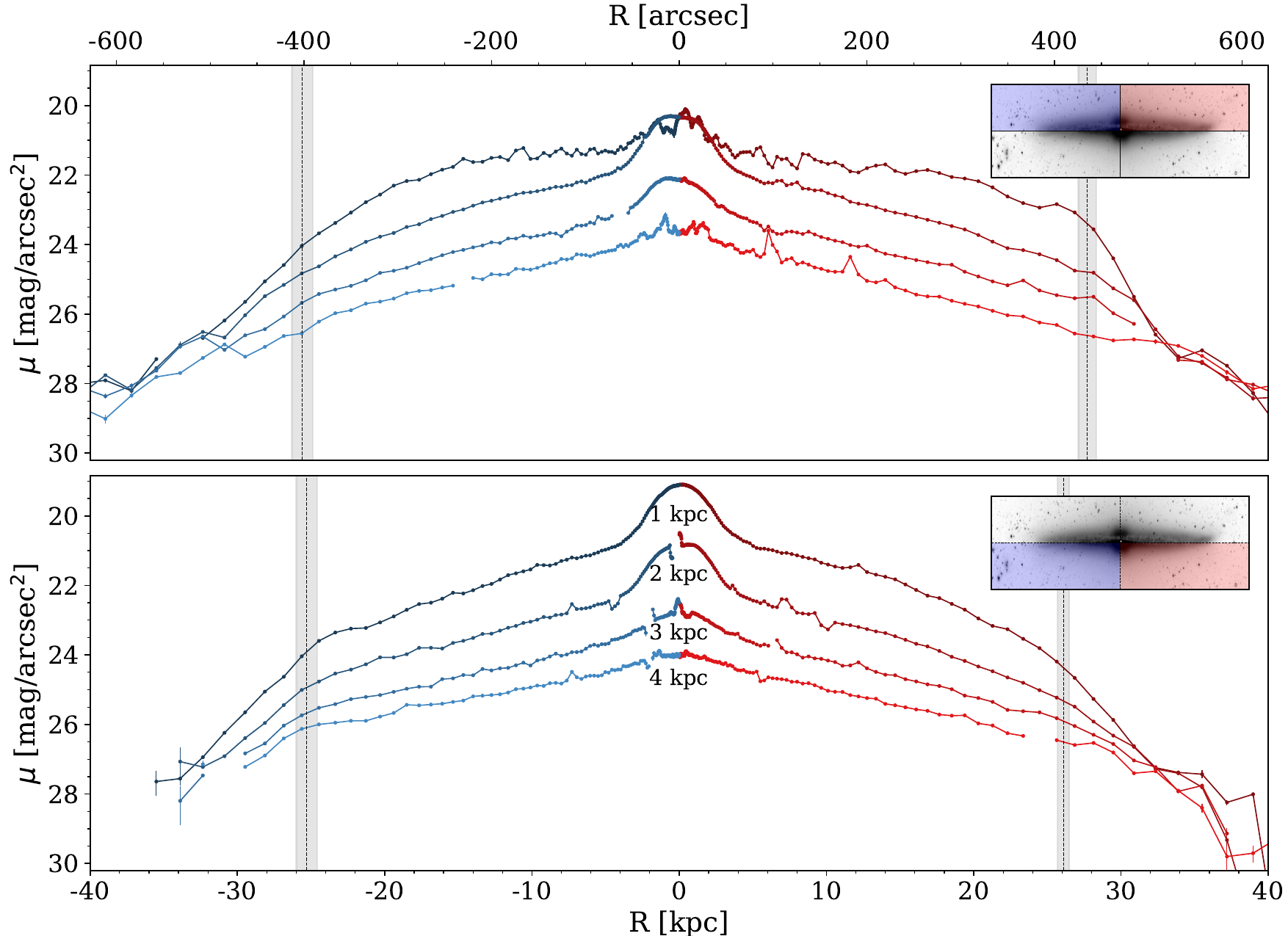}
\caption{ NGC~4565 radial surface brightness profiles in the $r$-band for the four disk quadrants and at four different heights above/below the mid-plane as indicated in the bottom panel (i.e. at 1, 2, 3, and 4~kpc height). The vertical dashed lines are placed at the mean truncation radius of each quadrant. The standard deviation of the truncation position distribution is represented with shaded grey regions. In each panel, the upper-right insets with the image of NGC~4565 show two colour-shaded regions that indicate the quadrants where the RSBPs were extracted and colour-coded accordingly. The vertical development of the truncation radius is sharper on the northwest side (right; in red) of the disk while the southeast side (left; in blue) shows two breaks along the radial distribution. The first break is at a radius of $\sim 17$~kpc while the second break corresponds to the truncation. Nonetheless, the truncation radius remains similar within the uncertainties for the three quadrants without a strong warp feature.}
\label{Fig:quadrants}
\end{figure*}

\begin{table*}
\centering
\caption{Scale lengths in kiloparsecs of the profile fits on both sides of the truncation for NGC~4565 in the ultra-deep $g$, $r$, and $i$ bands and for \ion{H}{i} data. For each height $\mid z \mid$ and each wavelength, we give the numerical value of the exponential fit before and after the truncation radius ($h_{\rm bef}$ and $h_{\rm aft}$).} \vspace{4mm}
\label{tabla:scalelenghts} 
\begin{tabular}{c|cccccccc}
\hline
  & \multicolumn{2}{c}{$g$} & \multicolumn{2}{c}{$r$} & \multicolumn{2}{c}{$i$} &
  \multicolumn{2}{c}{\ion{H}{i}} \\
		 $\mid z \mid$ & $h_{\rm bef}$ & $h_{\rm aft}$ & $h_{\rm bef}$  & $h_{\rm aft}$ & $h_{\rm bef}$ & $h_{\rm aft}$ & $h_{\rm bef}$ & $h_{\rm aft}$ \\
		\hline
0.0 & $8.6 \pm 0.3$ & $ 1.6 \pm 0.2$ & $7.3 \pm 0.3$ & $ 1.6 \pm 0.1$ & $5.9 \pm 0.2$ & $ 1.8 \pm 0.1$ & $17.7 \pm 0.7$ & $ 9.0 \pm 0.9$ \\

0.5 & $10.2 \pm 0.4$ & $1.6 \pm 0.2$ & $7.9 \pm 0.3$ & $ 1.7 \pm 0.2$ & $7.1 \pm 0.2$ & $ 1.7 \pm 0.2$ & $28.8 \pm 0.8$ & $ 11.3 \pm 1.1$  \\

1.0 & $9.1 \pm 0.1$ & $ 2.0 \pm 0.2$ & $7.1 \pm 0.2$ & $2.1 \pm 0.2$ & $6.7 \pm 0.3$ & $ 2.0 \pm 0.2$ & $-$ & $-$ \\

1.5 & $7.3 \pm 0.3$ & $ 2.7 \pm 0.2$ & $5.8 \pm 0.2$ & $ 3.0 \pm 0.3$ & $6.7 \pm 0.2$ & $ 2.5 \pm 0.2$ & $21.7 \pm 1.2$ & $ 7.0 \pm 1.4$ \\

2.0 & $7.4 \pm 0.3$ & $ 3.5 \pm 0.2$ & $7.0 \pm 0.2$ & $3.7 \pm 0.2$ & $7.2 \pm 0.2$ & $ 3.0 \pm 0.3$ & $-$ & $-$ \\

2.5 & $8.8 \pm 0.3$ & $ 3.9 \pm 0.3$ & $8.4 \pm 0.2$ & $ 4.1 \pm 0.2$ & $8.6 \pm 0.3$ & $3.3 \pm 0.2$ & $-$ & $-$ \\

3.0 & $10.1 \pm 0.3$ & $4.8 \pm 0.3$ & $10.0 \pm 0.2$ & $ 4.7 \pm 0.2$ & $9.5 \pm 0.2$ & $ 3.8 \pm 0.3$ & $-$ & $-$  \\

3.5 & $11.0 \pm 0.4$ & $6.0 \pm 0.3$ & $10.9 \pm 0.3$ & $ 5.8 \pm 0.3$ & $10.2 \pm 0.3$ & $ 4.9 \pm 0.3$ & $-$ & $-$ \\

4.0 & $11.6 \pm 0.4$ & $6.4 \pm 0.4$ & $11.5 \pm 0.3$ & $ 6.5 \pm 0.3$ & $11.2 \pm 0.4$ & $ 5.2 \pm 0.3$ & $-$ & $-$ \\
\end{tabular}
\end{table*}

The location of the radial position of the truncation remains constant up to 3~kpc, in agreement with \citetalias{MartinezLombilla2019a}. Further up, we have two more detections at 3.5 and 4~kpc but they are located at an inner radius (see Fig.~\ref{Fig:TruncPosBoth}). In other words, we see a shift in the radial location of the truncation towards the centre of the galaxy above 3~kpc. Then, the higher the altitude, the closer to the galactic centre. This shift is contributing to the differences with previous studies when obtaining the mean radial position of the truncation. In this work, the mean truncation radius in the optical broad bands is 25.9~$\pm$~0.4~kpc, a slightly smaller distance than the 26.4~$\pm$~0.4~kpc obtained by \citetalias{MartinezLombilla2019a} (see their Sect.~4.4), but in agreement within errors. However, if we do not take into account the two higher detections at 3.5 and 4~kpc in our estimation of the mean location of the truncation, we got a value of 26.1~$\pm$~0.2~kpc. This is a similar result to what was found before by \citetalias{MartinezLombilla2019a}.

Some other works have also found this independence on the location of the truncation with high above/below the galaxy mid-plane in edge-on disk galaxies \citep{deJong2007, Comeron2012, Comeron2017}. However, none of them have reached our combination of high spatial resolution, high altitude in the disk, and ultra-deep imaging. In the case of \citet{Comeron2012, Comeron2017}, they also studied the vertical light distribution of the disk of NGC~4565, between $\sim$2.7--7~kpc above the galaxy mid-plane. They found the truncation radius located at 27.5~$\pm$~0.8~kpc up to a high of 3~kpc using lower spatial resolution data. They did not detect any evidence of truncation at large heights.

We also explored whether there is any significant difference in the position of the truncation depending on the disk quadrant. To do that, we extracted RSBPs from each disk quadrant at five different heights above/below the mid-plane as shown in Fig.~\ref{Fig:quadrants}. The truncation is systematically sharper in the right side of the disk (NW; red profiles in Fig.~\ref{Fig:quadrants}). Also, the surface brightness of the truncation is slightly brighter ($\sim$0.3~mag~arcsec$^{-2}$) in the right disk, although this could be expected from the fact that NGC~4565 is not perfectly edge-on. The light profiles show different features depending on the disk side. On the right side (NW), the profiles are composed of two exponential light distributions that intersect at the truncation radius. On the left side of the disk (SE), the light shows an additional break located at $\sim 21$~kpc. This inner break is clearly visible  in the mid-plane profile but it gradually vanishes as height increases. The asymmetry in the disk of NGC~4565 has been previously reported by \cite{Naeslund1997, Wu2002} and \cite{Gilhuly2020}. The latter authors associated the disk asymmetry with a tidal ribbon produced by the fan-like feature located around the northwest disk side.

From the aforementioned profiles, we measured the truncation radius for each disk quadrant separately. Despite the disk light distribution asymmetry, the truncation radius remains at the same location and almost constant within the uncertainties for all the quadrants and heights above/below the mid-plane up to where we have enough S/N, with the only exception of the quadrant with a clear warp feature (north quadrant). This is shown in the central and bottom panels in Fig.~\ref{Fig:TruncPosBoth}. When there is no warp, the truncation radius remains constant. However, the truncation radius systematically increases with height above the galaxy mid-plane in the presence of the warp. It seems as if the truncation traces the warp feature, at least, up to 3~kpc height. The possible role of the warp on the origin of the disk of NGC~4565 will be further discussed in Sect.~\ref{sec:warpRole}.

In a previous work, \cite{Gilhuly2020} reported the truncation location at a smaller radius on the left (SE) side of the disk. However, we consider that the truncation at a smaller radius in the SE side of the disk is an inner break. Then, at $~26$~kpc, we measure a second decline in the light distribution that coincides with the truncation radius on the right side of the disk. The higher spatial resolution and depth of our data in comparison with that in \cite{Gilhuly2020} allowed us to distinguish both disk features.

\subsection{Truncation radius remains independent of the wavelength} \label{subsection:TrucWavelresults}

\begin{figure}
\centering
\includegraphics[width=\hsize]{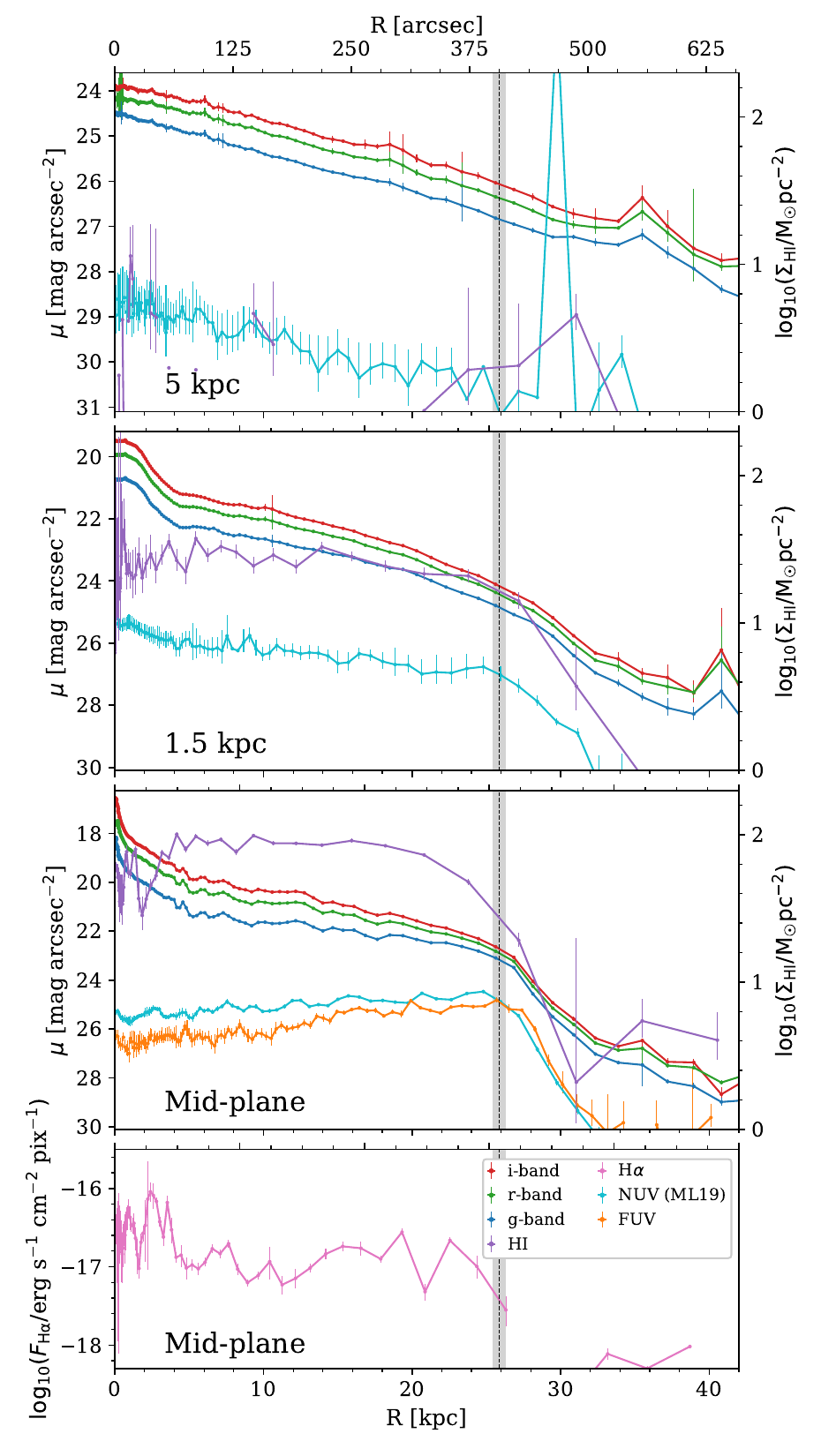}
  \caption{NGC~4565 radial surface brightness profiles in all the wavelength ranges used in this work, from FUV to \ion{H}{i}. The right y-axis is in surface mass density units to account for the \ion{H}{i} data. Each panel shows the RSBP at the indicated height above/below the mid-plane. In the cases of FUV and H$\alpha$ we only extracted profiles in the mid-plane (see Sect.~\ref{subsubsec:Halphaprof}). The NUV profiles are those from \citetalias{MartinezLombilla2019a}. The vertical dashed lines indicate the mean radius of the truncation in the broad-band optical data as obtained from the data points in the top panel of Fig~\ref{Fig:TruncPosBoth}. The shaded grey regions are the standard deviation of the truncation radius distribution. The truncation feature is present in all wavelengths at the same radius until it completely vanishes beyond 4~kpc height.
          }
     \label{Fig:ComparePanel}
\end{figure}

The radial position of the NGC~4565 truncation is the same, within errors, for our full wide wavelength range, independently of the height above/below NGC~4565 mid-plane. We find the same results when measuring the truncation radius in each disk quadrant separately. This is illustrated in Fig.~\ref{Fig:ComparePanel} and also in Fig.~\ref{Fig:TruncPosBoth} for the optical $g$, $r$, and $i$ bands only. For any given altitude, the radial location of the truncation is not systematically either closer or further from the galaxy centre in any of the filters. This result confirms our expectations as previous studies of NGC~4565 found this independence on the radial location of the truncation. \cite{Gilhuly2020} extracted RSBPs in $g$ and $r$ band with the same outcome, so as \citetalias{MartinezLombilla2019a}, in a wide wavelength range, from NUV to near-infrared ($3.6 \, \mu$m).

There is an evident sharp decline in the brightness of the H$\alpha$ emission around the truncation region in the mid-plane. We also detect such a decline in surface mass density in \ion{H}{i} gas component around the truncation that gradually vanishes with height to then completely disappears beyond 3~kpc above/below the galaxy mid-plane. We detect the radius of truncation feature in the \ion{H}{i} gas at $24.0 \pm 1.2$, $24.5 \pm 1.2$ and $27.4 \pm 1.6$~kpc for the mid-plane, 0.5 and 1.5~kpc height respectively (see the whole set of \ion{H}{i} RSBPs in Fig.~\ref{Fig:Hi-panel}). Similar trend is shown in both FUV and NUV data (see Fig.~\ref{Fig:ComparePanel}). It is clear that the disk truncation in the FUV, H$\alpha$ and \ion{H}{i} data is also detected although it is affected by the lower spatial resolution and/or depth (as previously explained in Sects.~\ref{subsection:UV}, \ref{subsubsec:Halphaprof} and \ref{subsubsec:HIprof}). These findings extend the independence of the location of the truncation in the Milky Way-like galaxy NGC~4565 towards a wider wavelength range, from FUV to \ion{H}{i}, in all measured wavelengths (i.e., FUV, NUV, $g$, $r$, $i$, H$\alpha$, $3.6 \, \mu$m, and \ion{H}{i}).

Our findings are in good agreement with \cite{DiazGarcia2022}. They extracted light profiles of the disk of UGC~7321 from FUV, optical ($grz$ combined mean), and NIR ($3.6 \, \mu$m) images. They reported the discovery of a truncation at and above the mid-plane at all the probed wavelength ranges. This is particularly interesting due to the pristine nature of UGC~7321, a well-studied edge-on, low-mass, diffuse, isolated, bulgeless, and ultra-thin galaxy. As suggested by \cite{DiazGarcia2022}, this supports that disks and truncations can form via internal mechanisms alone.

\subsection{PSF affects the outskirts of the disk of NGC~4565 but not the truncation radius} \label{subsection:PSFresults}

To ensure that our truncation measurements were not affected by the PSF, we fitted analytical 2D models convolved with the PSF to the ultra-deep optical broad-band data (see Sect.~\ref{subsection:NGC_PSFcorr}). From these models, we built the galaxy PSF deconvolved models, which are an approximation to the observed images, but without the PSF effects and with the noise and intrinsic asymmetries of the galaxy such as the warp. We refer to those models as the ``galaxy models''.

The PSF affects the light distribution in the outermost regions of the disk of NGC~4565 as there is more flux in the observed data than in the galaxy models as it has also previously reported in \citetalias{MartinezLombilla2019a}. Despite that, the truncation radius remains the same within the uncertainties when accounting for the PSF.

To prove the above, we extracted a full set of RSBPs from the final models of NGC~4565 in the optical broad-bands (Figs.~\ref{Fig:g-model-panel}, \ref{Fig:r-model-panel} and \ref{Fig:i-model-panel}). From these profiles, we determined the radial positions of the truncation at each height above/below the NGC~4565 mid-plane as described in Sect.~\ref{subsection:TruncationPosition}. We did the same for the corresponding RSBPs extracted from the observed data from Figs.~\ref{Fig:g-panel}, \ref{Fig:r-panel} and \ref{Fig:i-panel}. We show all those radial positions of the truncation in the top panel of Fig.~\ref{Fig:TruncPosBoth}. As previously stated, there are no systematic trends or behaviour  between both sets of data.

Our galaxy model, shown in Fig.~\ref{Fig:model-profile}, is able to properly reproduce and recover the light in the outskirts of the galaxy up to $\sim$~1000~arcsec (i.e. $\sim$~60~kpc). This is shown in the profiles and also in the images below, where the combination of the four galaxy components (i.e. bulge, bar, disk and halo) successfully mimics the shape of NGC~4565. The residuals basically account for the dusty regions surrounding the galaxy mid-plane and some asymmetries such as the disk warp, both impossible to reproduce with the available analytical functions in the software (see details in Sect.~\ref{subsection:NGC_PSFcorr}). \citet{Gilhuly2020} also obtained a model of NGC~4565 but using just two components: a 3D broken exponential disk and a 2D S{\'e}rsic function. By comparing both approaches, their model underestimates the light in the outer parts of the disk by up to a 21~$\%$ in $g$-band \citep[][see caveats of the model in their Sect.~4.2]{Gilhuly2020}. Nevertheless, both works agree in the fact that there is no significant variation in the radial location of the truncation between the observed data and the models \cite[note that][extracted an azimuthally averaged profile, that is, from elliptical isophotes]{Gilhuly2020}.

Regarding the shape of the RSBPs in the truncation region, the truncation always shows the sharpest decline in the mid-plane of the galaxy and becomes less prominent at large heights in both the galaxy models and the observed data. However, the truncation feature remains clear at higher altitudes above/below mid-plane in the models than in the observed data, for the three bands. This is showed in detail in the figures of the Appendix~\ref{appendix:RSBP} (Figs.~\ref{Fig:g-panel}, \ref{Fig:r-panel}, \ref{Fig:i-panel}, \ref{Fig:g-model-panel}, \ref{Fig:r-model-panel} and \ref{Fig:i-model-panel}).

\section{Star formation activity at the edge of the disk} \label{section:SFRresults}

\begin{figure}
\centering
\includegraphics[width=\hsize]{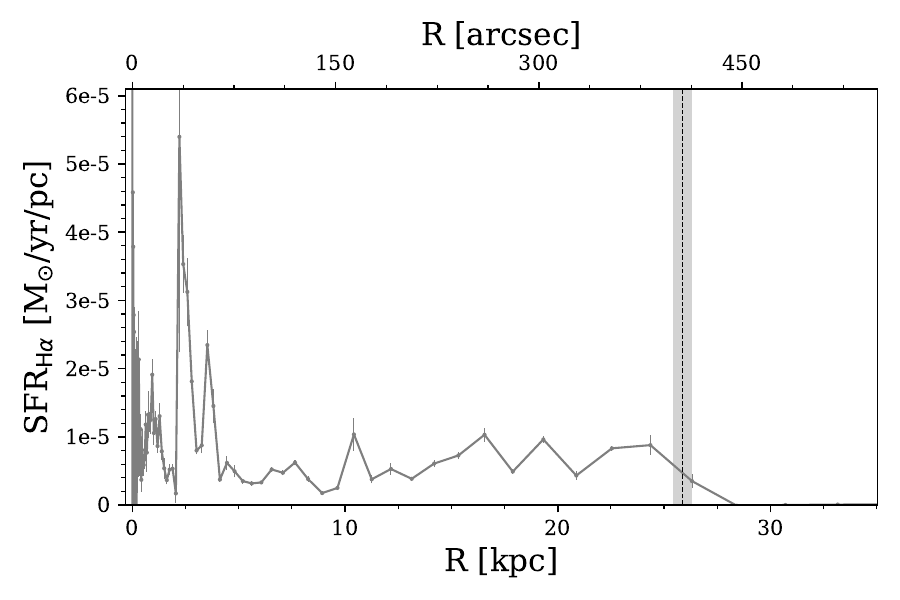}
  \caption{SFR per radial parsec as a function of the disk radius in NGC~4565 obtained from the H$\alpha$ data. The vertical dashed line indicates the mean radius of the truncation in the broad-band optical data at 25.9~$\pm$~0.4~kpc, and the shaded grey region is the uncertainty of the mean estimation. The radial bins are logarithmically spaced so the innermost bins are smaller that those in the outer parts of the disk. For the first time, we measure that NGC~4565 is forming stars at the truncation radius at a rate of $3.5 \times 10^{-6} \pm 5 \times 10^{-6}$ M$_{\odot}$/yr/pc, or a total of $0.011 \pm 0.002$ M$_{\odot}$/yr within the bin.
  }
     \label{Fig:SFR}
\end{figure}

We found, for the first time, that quiescent NGC~4565 galaxy is forming $0.011 \pm 0.002$ solar masses of stars per year at the truncation radius, at a rate of $3.5 \times 10^{-6} \pm 5 \times 10^{-6}$ M$_{\odot}$/yr/pc. The SFR values per parsec at the galaxy mid-plane as a function of the radial distance are shown in Fig.~\ref{Fig:SFR} as a radial SFR profile extracted from the continuum-subtracted H$\alpha$ imaging (see Sect.~\ref{subsection:SFR}). This allows us to investigate the star formation activity in the vicinity and beyond the truncation radius of NGC~4565.

We verified the drop-off in SFR at the disk truncation. Right before the truncation and three more times throughout the galaxy disk, there are peaks of SFR with values of $\sim 1 \times 10^{-5}$ M$_{\odot}$/yr/pc, located at radii  10.4, 16.6, 19.3 and 24.4~kpc. After the truncation, the SFR decreases until it is completely quenched at radius of $\sim$~28~kpc.

The total SFR in the disk is 0.63~$\pm$~0.01~M$_{\odot}$~yr$^{-1}$ while the median SFR is 0.02~$\pm$~0.01~M$_{\odot}$~yr$^{-1}$. To get these values we avoided the more active inner 4~kpc in radius that is also the region where we could not properly subtract the sky emission and the continuum due to light gradients in the observed data so the flux is overestimated. NGC~4565 is a widely studied quiescent nearby late-type spiral galaxy with both a low SFR and SFR surface density. In comparison with previous works, \citet{Heald2012} determined a total SFR of 0.67~$\pm$~0.10~M$_{\odot}$~yr$^{-1}$ (among the lowest in their HALOGAS sample), while \citet{Wiegert2015} obtained a value of 0.73~$\pm$~0.02~M$_{\odot}$~yr$^{-1}$. These total SFR values are smaller to ours if we consider the whole disk. However, our measurements in the innermost disk region are probably overestimated and, on the other side, \citet{Heald2012} and \citet{Wiegert2015} considered smaller distances to NGC~4565 than in our work.

\begin{figure*}
\begin{center}
\includegraphics[width=\hsize]{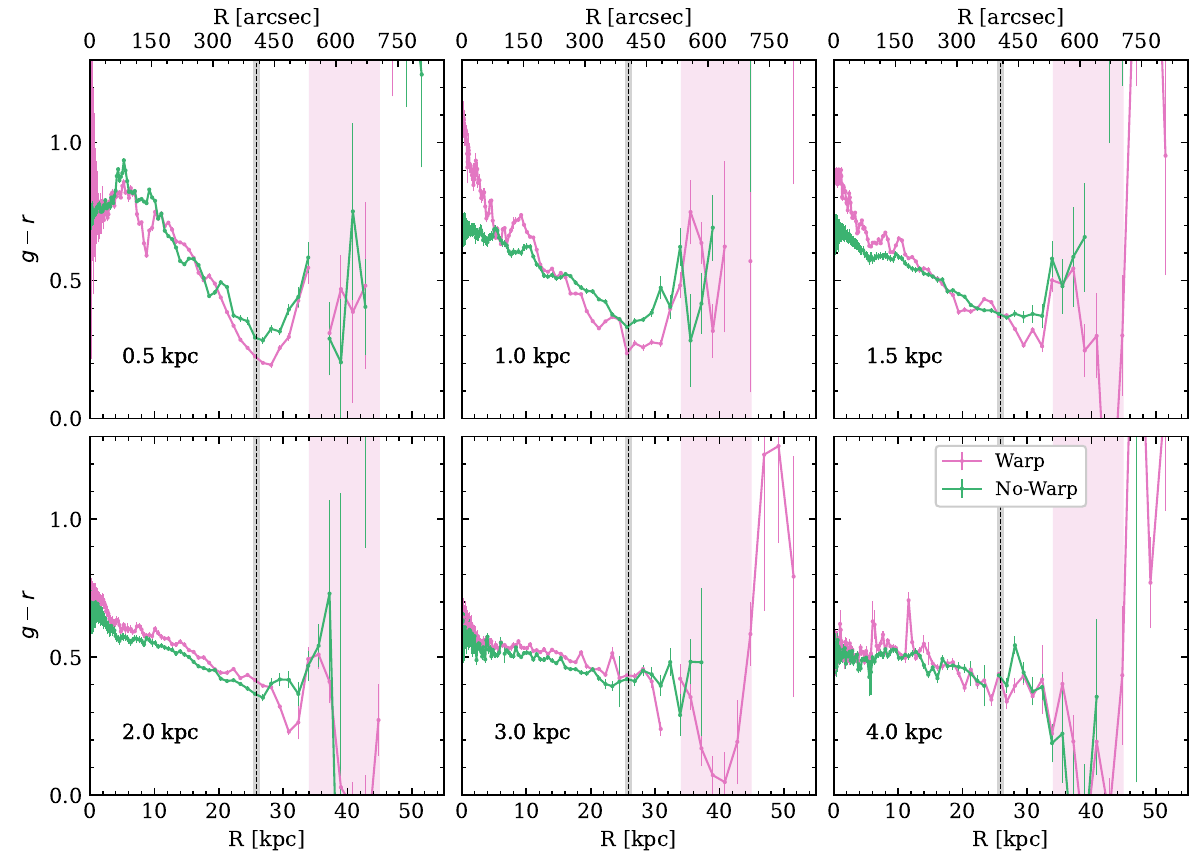}
\caption{$g$-$r$ colour radial profile at six different heights above/below the NGC~4565 mid-plane up to 4~kpc but excluding the mid-plane. Solid pink lines are the profiles from the disk region affected by a warp (north quadrant) while the green ones indicate the profiles of the region of the disk with no warp feature (west quadrant). The vertical shaded pink area highlights the warp region. The colour profiles show two moving U-shape features, characteristic of a star formation threshold. The first U-shape is at the truncation radius and close to the mid-plane, while the second traces the warp beyond $R \sim 38$~kpc and above 1.5~kpc height. }
\label{Fig:g-r-color}
\end{center}
\end{figure*}

We detect a clear characteristic U-shaped age gradient up to 2~kpc in height around the location of the truncation in all the optical and NUV colour profiles allowed by our data. The $g$-$r$ colour radial profiles (Fig.~\ref{Fig:g-r-color}) show the most prominent U-shape feature in optical wavelength allowing also for high S/N beyond the truncation. The U-shape is clearly detected up to 1~kpc in height and then slowly disappears when moving to larger heights. In Fig.~\ref{Fig:g-r-color}, we show a colour profile for the region of the disk with a warp (the upper-right disk; red region in the inset of the top panel in Fig.~\ref{Fig:quadrants}) and a region less affected by the warp (the lower-right disk; red region in the inset of the bottom panel in Fig.~\ref{Fig:quadrants}). The potential role of the warp in the truncation properties will be further discussed in Sect.~\ref{sec:warpRole}. 

The bluest colour (i.e., the youngest stellar population) is 0.23~$\pm$~0.02 at a radius of 27~kpc in the $g$-$r$ colour radial profiles. Beyond the truncation, the colour reddens again. In the case of the $g$-$i$ colour profile, a less prominent minimum of age is located at a radius of 27~kpc with a colour value of 0.43~$\pm$~0.02. Finally, in the $r$-$i$ colour, the U-shape feature shows just a hint of its presence in the profiles extracted in the galactic plane and at 0.5~kpc high. Again, we found the minimum colour value (i.e., age) of 0.20~$\pm$~0.02 at a radius of 27~kpc.

It is important to note that the mid-plane regions of the disk of NGC~4565 are strongly affected by dust which could influence our measured colours. However, by comparing the light distribution in NUV, optical, and NIR regimes, \citetalias{MartinezLombilla2019a} found that extinction is mostly limited to the inner areas and there is almost no dust extinction at the location of the truncation and beyond.

This U-shape feature in the colour profiles has originally been found in observations of type II breaks \citep[see e.g.][]{Azzollini2008a, Bakos2008} at lower radial distances. However, recent works have related the U-shape with the truncation location. The first detection of this feature was by \citetalias{MartinezLombilla2019a}, who found it in the Milky Way-like galaxies NGC~5907 and NGC~4565 \citep[for this galaxy later confirmed by][]{Gilhuly2020}. \cite{DiazGarcia2022} observed the U-shape at the truncation location of the low-mass ultra-thin galaxy UGC~7321 for both the thin and thick disc. In addition, \cite{Chamba2022} measured a characteristic U-shape from $g$-$r$ colour profiles in a broad variety of galaxy types (dwarfs to ellipticals) and stellar masses ($10^{7}$~M$_{\sun} < M_{*} < 10^{12}$~M$_{\sun}$).

Theoretical studies on the origin of type II breaks in galaxy disks by \citet{Rovskar2008b} predict that if breaks and truncations share the same formation process, this would be the result of a radially declining SFR. In that scenario, the minimum stellar age value would be at the location of the truncation or break, combined with radial migration further out, producing the U-shape in the radial colour profiles. As reported above, we find a U-shape in our colour profiles (see Fig.~\ref{Fig:g-r-color}). Also, these theoretical predictions are in good agreement with our results shown in Fig.~\ref{Fig:SFR}, where the SFR decreases just before reaching the truncation radius.

The star formation threshold linked to the U-shape in the colour profiles (Fig.~\ref{Fig:g-r-color}) traces, on the one hand, the youngest stellar population at the truncation location and, on the other hand, a likely radial migration of stars towards the outer regions associated with the redder colours beyond the truncation radius \citep[see related discussion in \citetalias{MartinezLombilla2019a} and][]{DiazGarcia2022}. The lack of H$\alpha$ emission beyond the truncation (see Fig.~\ref{Fig:GalIm}) supports this scenario as it is indicative of a death of recent star formation ($< 10$~Myr). However, there is FUV emission beyond the truncation, tracing past star formation of several tens to 100~Myr ago \citep{Kennicutt1998}. The NUV flux traces a slightly older population of stars formed $\lesssim 300$~Myr ago \citep{Kennicutt1998} and reaches higher altitudes above/below the galaxy mid-plane than both H$\alpha$ and FUV. The NUV emission extents up to similar radial distances beyond the truncation than the FUV. We suggest the outward radial migration of stars from the star-forming threshold at the truncation radius to the outer disk as the main mechanism populating the external disk. The young stars then become visible in FUV and NUV wavelengths. Thus, the further from the truncation radius, the older the stellar population. In summary, our observational measurements are consistent with a common origin for breaks and truncations.

\section{Vertical extent of the disk and its origin} \label{Sec:vertDisk}

In \citetalias{MartinezLombilla2019a}, we found that at the truncation could be observed up to a height of $\sim 2.9$~kpc. This agrees with our results of a coherent disk structure up to 3~kpc as the truncation is located at a constant radius of 26.1~$\pm$~0.2~kpc. However, the extra truncation detections in this work that extend this feature up to 4~kpc from the mid-plane suggest different possible scenarios for the disk vertical structure.

If the truncation in the mid-plane is a phenomenon associated to the end of the star formation in the disk and the truncation is moving with time to further distances, then one should expect the radial position of the truncations above/below the mid-plane to be located at smaller radial distances. This is supported by the fact that stars in galactic disks are subject to migrations in both the radial and vertical directions \citep[see for reviews][]{Sellwood2014, Debattista2017}. Our highest truncation detections at 3.5 and 4~kpc height, which are at closer radial distances to the centre of NGC~4565 (see top panel in Fig.~\ref{Fig:TruncPosBoth}), support this inside-out disk growth scenario.

Assuming a similar physical origin for the truncation in both the galactic plane and at higher altitudes \citep[e.g.,][]{deJong2007, RadburnSmith2012}, the brightness of a thick disk component is still not enough to outshine the truncation feature from the thin disk at $\sim 3-4$~kpc height. There is also the possibility of a flared thin disk that dominates at large heights in the outskirts of the galaxy. The ($g -r$) colour map in Fig.~\ref{Fig:ColorMaps} shows evidence of a flare of the younger (bluer) stellar populations in the outer part of the thin disk ($R>15$~kpc). This flaring is more clear in the SE side of the disk (left), where the warp is less prominent. Also, as NGC~4565 is not perfectly edge-on ($i = 88.5 \pm 0.5$~deg; \citetalias{MartinezLombilla2019a}), the flare could be partially blurred by the addition of light along the line of sight.

Numerical simulations suggest that flaring cannot be avoided due to a range of different dynamical effects. The main disk flaring mechanism is probably satellite–disk interactions \citep[e.g.,][]{Kazantzidis2008, Villalobos2008, Bournaud2009, Qu2011}, but mergers \citep{Abadi2003}, misaligned gas infall \citep{Scannapieco2009, Roskar2010} and reorientation of the disk rotation axis \citep{Aumer2013} can produce disk flaring too. However, simulations by \cite{Minchev2015} led them to propose that purely secular evolution, based in an inside-out growing disks, also cause flared disks. The latter scenario satisfies both our observational constraints and the environment conditions in NGC~4565.

From observations, \cite{Iodice2019b} found flared thin disks in three S0 edge-on galaxies within a dense environment in the Fornax cluster from isophotal and colour analysis. \cite{Pinna2019b, Pinna2019a} supported these findings with high-quality MUSE stellar kinematic data. \cite{Kasparova2020}, using deep long-slit spectroscopic data across the massive edge-on galaxy NGC~7572, measured a flare in both the thin and thick disks with similar radial disk scales. However, is important to consider that all these galaxies are located in fairly dense environments, unlike the isolated Milky Way-like galaxy NGC~4565, potentially affecting the formation process of the flare. In the Milky Way, several authors have also reported flaring of younger stellar populations \citep[e.g.,][]{Feast2014, Kalberla2014, Carraro2015a, Ding2021, Chrobakova2022}.

In terms of the stellar population ages in the NGC~4565 disk, we see that the ($g-r$) colour is clearly below 0.8 at the truncation/warp, while ($g-r$)~$\sim 1$ in the innermost parts (Fig.~\ref{Fig:ColorMaps}), reflecting a young population. The individual light distribution of the galaxy in each wavelength (Fig.~\ref{Fig:GalIm}) highlights a thin H$\alpha$ disk of stars born $\sim 10$~Myr ago. The FUV data exhibit a rather thicker thin disk of $\sim 2$~kpc (accounting for inclination effects), tracing star formation that occurred from several tens to $\sim 100$~Myr ago. The thin disk is even thicker in NUV data, which traces $<300$~Myr populations \citep{Kennicutt1998}. However, we do not see star formation in the thick disk. The \ion{H}{i} gas is also concentrated within the thin disk area.

All these findings suggest that the disk of the Milky Way-like galaxy NGC~4565 has been formed mainly through internal processes. In particular, the thin disk stars would be born in the galaxy mid-plane (up to $\sim 1$~kpc) and then migrate along the vertical axis, supporting an inside-out and down-up growth of the disk with an outer flare feature.

\section{The role of the disk warp of NGC~4565 in the truncation radius} \label{sec:warpRole}

\begin{figure*}
\centering
\includegraphics[width=\textwidth]{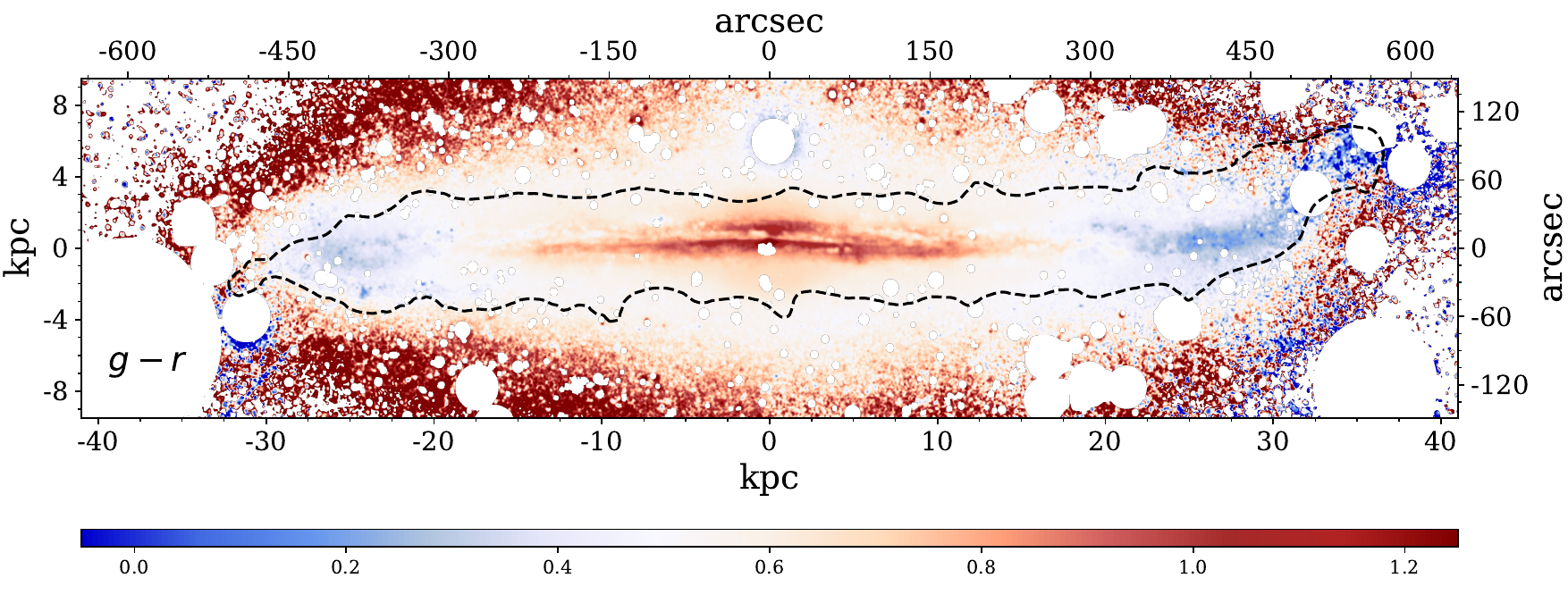}
  \caption{$g-r$ colour map of NGC~4565 with values indicated in the bottom colour bar. The dashed black line shows the outermost \ion{H}{i} isocontour. The warp is traced with bluer colours in the off-plane regions along the upper-right side of the disk. We can also see the bluer and little dust-affected regions around the truncation ($R \sim 26$~kpc). Also, the \ion{H}{i} emission follows the optical light distribution.}
     \label{Fig:ColorMaps}
\end{figure*}

We clearly see a warp in the disk of NGC~4565 (Fig.~\ref{Fig:GalIm}) and we report, for the first time, the warp detection in FUV and NUV data. This warp in NGC~4565 has been previously reported in optical and \ion{H}{i} data \citep{vanderKruit1979, Rupen1991, Naeslund1997, Radburn-Smith2014, Yim2014, MartinezLombilla2019a, Gilhuly2020}. However, our ultra-deep data indicate a very prominent warp in the north side (upper-right disk quadrant) with $\sim 6-7$~kpc length, while it is fainter and shorter in the south edge. As we do not see the warp in H$\alpha$, this indicates that there is recent ($\gtrsim 100$~Myr ago) but not ongoing star formation in this disk structure. The warp also shows clear \ion{H}{i} gas content. These measurements are in agreement with the stellar populations of less than 600~Myr old reported for the warp by \cite{Radburn-Smith2014}.

The presence of the warp affects the truncation feature (see Sect.~\ref{subsection:TrucPosresults}), contrary to our findings in  \citetalias{MartinezLombilla2019a}. The extraordinary depth of the new optical data allows us to measure a truncation radius that is systematically larger the higher the detection above the NGC~4565 mid-plane (see central panel in Fig.~\ref{Fig:TruncPosBoth}). Both the ($g-r$) colour profiles in Fig.~\ref{Fig:g-r-color} and the maps in Fig.~\ref{Fig:ColorMaps}, highlight younger (bluer) stellar populations at the truncation radius in the presence of the prominent warp. The \ion{H}{i} map also reveals that the truncation radii is related to the warp location. From Frig.~\ref{Fig:GalIm} we can estimate the onset of the warp in \ion{H}{i} at $\sim 26-27$~kpc, coinciding in radius with the mean radial truncation value in the optical bands  (25.9~$\pm$~0.4~kpc).

We measured a two moving U-shapes in the colour profiles (see Fig.~\ref{Fig:g-r-color}) of the disk quadrant with a warp. Such U-shapes are indicative of a threshold in star formation. We find a first U-shape tracing the truncation at $R \sim 26$~kpc (see Sect.~\ref{section:SFRresults}) and a second one further outside and at higher altitudes off the galaxy mid-plane following the warp beyond $R \sim 38$~kpc and above 1.5~kpc height. To our knowledge, this is the first time such a feature has been detected in a disk warp.

Truncations have previously been linked to the maximum angular momentum of the protogalactic cloud \citep[e.g.,][]{vanderkruit1987a, Martin-Navarro2012}, but also to the presence of disk warps \citep{vanderKruit1979, Naeslund1997} and to star forming thresholds \citep[][\citetalias{MartinezLombilla2019a}]{Kennicutt1989}. We find that the truncation of the Milky Way-like galaxy NGC~4565 is linked to two of the above-mentioned mechanisms: a threshold in star formation and the warp.

\section{Conclusions} \label{section:conclusions}

We perform a multi-wavelength vertical analysis of the disk of the edge-on Milky Way-like galaxy NGC~4565 to establish the height up to where the truncation is still present at the edge of the disk and how is it related with disk properties such as the warp or the disk thickness. This is a follow-up work of \citetalias{MartinezLombilla2019a} including a wider wavelength range and ultra-deep optical data from the 4.2\,m WHT. Our findings can be summarised in the following statements:

\begin{enumerate}
    \item We obtain a mean radial truncation position of 25.9~$\pm$~0.4~kpc for the $g$, $r$, and $i$ bands up to a height of 4~kpc above/below the galaxy mid-plane. 

    \item  We confidently detect the truncation up to 4~kpc in the new ultra-deep optical images, that is, at locations 1~kpc higher than in any previous work for any studied galaxy. The location of the truncation radius remains constant up to a height of 3~kpc, in agreement with \citetalias{MartinezLombilla2019a}. Higher up, we report two more detections at 3.5 and 4~kpc but they are located at an inner radius.

    \item The truncation is systematically sharper on the right side of the disk (NW). Despite this, the truncation radius remains almost constant within the uncertainties for all the disk quadrants and heights above/below the mid-plane with the exception of the north quadrant, that is, the one with a clear warp.

    \item The radial position of the truncation is the same, within uncertainties, for our full wide wavelength range, independently of the height above/below NGC~4565 mid-plane. Combining these results with those in \citetalias{MartinezLombilla2019a}, we confirm that the truncation radius is similar at all measured wavelength ranges from FUV to \ion{H}{i}.

    \item Despite the effect of PSF scattering in the light distribution of NGC~4565, the radial position of the truncation remains unaffected (within uncertainties).

    \item Our observational measurements on the colour profiles (U-shape feature), SFR profile, and UV emission in the outer disk are in agreement with the predicted common origin for breaks and truncations. 

    \item We propose that the disk of NGC~4565 has been mainly formed through internal processes. The stars were born in the galaxy mid-plane (up to $\sim 1$~kpc) and then migrated vertically forming an outer flare. This supports an inside-out growth mechanism of the disk.

    \item We report the first detection of the warp of NGC~4565 in FUV and NUV data.

    \item The onset radius of the disk warp in \ion{H}{i} is located at $\sim 26-27$~kpc, coinciding in radius with the truncation location in the optical bands.

    \item Our findings suggest that the truncation of the Milky Way-like galaxy NGC~4565 is linked to both a threshold in star formation and the presence of a warp.

\end{enumerate}

This work highlights the connection between the truncation of the disk, the star formation activity, and the onset of the warp of the nearby galaxy NGC~4565 using an unprecedentedly wide wavelength range and ultra-deep optical data. Further analyses of large samples of edge-on galaxies, including studies of the dependence of these connections with the redshift, are required to understand the role of the truncations in galaxy disks.

\begin{acknowledgements}
We acknowledge constructive remarks by an anonymous referee that help to improve this paper. We thank Kijeong Yim for kindly sharing the VLA \ion{H}{i} reduced integrated maps with. CML thanks Sarah Brough and Sim{\'o}n D{\'i}az-Garc{\'i}a for useful discussions and support during the project. 
CML acknowledges the support of the Australian Research Council Discovery Project DP190101943. Parts of this research were supported by the Australian Research Council Centre of Excellence for All Sky Astrophysics in 3 Dimensions (ASTRO 3D), through project number CE170100013.
RIS acknowledges the funding by the Governments of Spain and Arag{\'o}n through the Fondo de Inversiones de Teruel; and the Spanish Ministry of Science, Innovation and Universities (MCIU/AEI/FEDER, UE) with grant PGC2018-097585-B-C21. 
We acknowledge financial support from the State Research Agency (AEI-MCINN) of the Spanish Ministry of Science and Innovation under the grants PID2019-107427GB-C32 and "The structure and evolution of galaxies and their central regions" with reference PID2019-105602GB-I00/10.13039/501100011033, from the ACIISI, Consejer\'{i}a de Econom\'{i}a, Conocimiento y Empleo del Gobierno de Canarias and the European Regional Development Fund (ERDF) under a grant with reference PROID2021010044, and from IAC projects P/300624 and P/300724, financed by the Ministry of Science and Innovation, through the State Budget and by the Canary Islands Department of Economy, Knowledge and Employment, through the Regional Budget of the Autonomous Community.
SC acknowledges support from the Ram\'on y Cajal programme funded by the Spanish Government (references RYC2020-030480-I), and from the State Research Agency (AEI-MCINN) of the Spanish Ministry of Science and Innovation under the grant “Thick discs, relics of the infancy of galaxies" with reference PID2020-113213GA-I00.
This research includes computations using the computational cluster Katana supported by Research Technology Services at UNSW Sydney. This research has made use of the SVO Filter Profile Service (\url{http://svo2.cab.inta-csic.es/theory/fps/}) supported by the Spanish MINECO through grant AYA2017-84089. This article is based on observations made in the Observatorios de Canarias del IAC with the WHT and INT operated on the island of La Palma by the Isaac Newton Group in the Observatorio del Roque de los Muchachos.
A.B. was supported by an appointment to the NASA Postdoctoral Program at the NASA Ames Research Center, administered by Oak Ridge Associated Universities under contract with NASA. A.B. is supported by a NASA Astrophysics Data Analysis grant (22-ADAP22-0118), program Hubble Archival Research project AR 17041, and Chandra Archival Research project ID \#24610329, provided by NASA through a grant from the Space Telescope Science Institute and the Center for Astrophysics Harvard \& Smithsonian, operated by the Association of Universities for Research in Astronomy, Inc., under NASA contract NAS 5-03127.

\newline
\newline

\textit{Facilities:} William Herschel Telescope (WHT), Isaac Newton Telescope (INT), Very Large Array (VLA), and Galaxy Evolution Explorer (GALEX), UNSW Katana cluster (\url{https://doi.org/10.26190/669x-a286}).  

\newline
\newline

\textit{Software:} \texttt{Astropy} (The Astropy Collaboration et al. 2018), \texttt{Photutils} v0.7.2 \citep{Bradley2020}, \texttt{Numpy} \citep{Oliphant2006}, \texttt{NoiseChisel} \citep{Akhlaghi2015, Akhlaghi2019}, \texttt{SExtractor} v2.19.5 \citep{Bertin1996}, \texttt{SWarp} v2.38.0 \citep{Bertin2002}, \texttt{SCAMP} v2.0.4 \citep{Bertin2006}, \textsc{imfit} \citep{Erwin2015}, \texttt{SAOImageDS9} \citep{Joye2003}.

\end{acknowledgements}

%

\begin{thebibliography}{122}
	\expandafter\ifx\csname natexlab\endcsname\relax\def\natexlab#1{#1}\fi
	
	\bibitem[{Abadi {et~al.}(2003)Abadi, Navarro, Steinmetz, \& Eke}]{Abadi2003}
	Abadi, M.~G., Navarro, J.~F., Steinmetz, M., \& Eke, V.~R. 2003, \apj, 597, 21
	
	\bibitem[{Akhlaghi(2019)}]{Akhlaghi2019}
	Akhlaghi, M. 2019, arXiv e-prints, arXiv:1909.11230
	
	\bibitem[{Akhlaghi \& Ichikawa(2015)}]{Akhlaghi2015}
	Akhlaghi, M. \& Ichikawa, T. 2015, \apjs, 220, 1
	
	\bibitem[{Alam {et~al.}(2015)Alam, Albareti, Prieto, Anders, Anderson,
		Anderton, Andrews, Armengaud, Aubourg, Bailey, Basu, Bautista, Beaton, Beers,
		Bender, Berlind, Beutler, Bhardwaj, Bird, Bizyaev, Blake, Blanton, Blomqvist,
		Bochanski, Bolton, Bovy, Bradley, Brandt, Brauer, Brinkmann, Brown,
		Brownstein, Burden, Burtin, Busca, \& Cai}]{Alam2015}
	Alam, S., Albareti, F.~D., Prieto, C.~A., {et~al.} 2015, \apjs, 219, 12
	
	\bibitem[{Athanassoula {et~al.}(1990)Athanassoula, Morin, Wozniak, Puy, Pierce,
		Lombard, \& Bosma}]{Athanassoula1990}
	Athanassoula, E., Morin, S., Wozniak, H., {et~al.} 1990, \mnras, 245, 130
	
	\bibitem[{Aumer \& White(2013)}]{Aumer2013}
	Aumer, M. \& White, S. D.~M. 2013, \mnras, 428, 1055
	
	\bibitem[{Azzollini {et~al.}(2008)Azzollini, Trujillo, \&
		Beckman}]{Azzollini2008a}
	Azzollini, R., Trujillo, I., \& Beckman, J.~E. 2008, \apjl, 679, L69
	
	\bibitem[{Bakos {et~al.}(2008)Bakos, Trujillo, \& Pohlen}]{Bakos2008}
	Bakos, J., Trujillo, I., \& Pohlen, M. 2008, \apjl, 683, L103
	
	\bibitem[{Bertin(2006)}]{Bertin2006}
	Bertin, E. 2006, in Astronomical Society of the Pacific Conference Series, Vol.
	351, Astronomical Data Analysis Software and Systems XV, ed. C.~{Gabriel},
	C.~{Arviset}, D.~{Ponz}, \& S.~{Enrique}, 112
	
	\bibitem[{{Bertin} \& {Arnouts}(1996)}]{Bertin1996}
	{Bertin}, E. \& {Arnouts}, S. 1996, \aaps, 117, 393
	
	\bibitem[{Bertin {et~al.}(2002)Bertin, Mellier, Radovich, Missonnier, Didelon,
		\& Morin}]{Bertin2002}
	Bertin, E., Mellier, Y., Radovich, M., {et~al.} 2002, in Astronomical Society
	of the Pacific Conference Series, Vol. 281, Astronomical Data Analysis
	Software and Systems XI, ed. D.~A. {Bohlender}, D.~{Durand}, \& T.~H.
	{Handley}, 228
	
	\bibitem[{Bland-Hawthorn \& Gerhard(2016)}]{BlandHawthorn2016}
	Bland-Hawthorn, J. \& Gerhard, O. 2016, \araa, 54, 529
	
	\bibitem[{Borlaff {et~al.}(2017)Borlaff, Eliche-Moral, Beckman, Ciambur,
		P{\'e}rez-Gonz{\'a}lez, Barro, Cava, \& Cardiel}]{Borlaff2017}
	Borlaff, A., Eliche-Moral, M.~C., Beckman, J.~E., {et~al.} 2017, \aap, 604,
	A119
	
	\bibitem[{{Borlaff} {et~al.}(2019){Borlaff}, {Trujillo}, {Rom{\'a}n},
		{Beckman}, {Eliche-Moral}, {Infante-S{\'a}inz}, {Lumbreras-Calle}, {de
			Almagro}, {G{\'o}mez-Guijarro}, {Cebri{\'a}n}, {Dorta}, {Cardiel},
		{Akhlaghi}, \& {Mart{\'{\i}}nez-Lombilla}}]{Borlaff2019}
	{Borlaff}, A., {Trujillo}, I., {Rom{\'a}n}, J., {et~al.} 2019, \aap, 621, A133
	
	\bibitem[{Bournaud {et~al.}(2009)Bournaud, Elmegreen, \& Martig}]{Bournaud2009}
	Bournaud, F., Elmegreen, B.~G., \& Martig, M. 2009, \apj, 707, L1
	
	\bibitem[{Bradley {et~al.}(2020)Bradley, Sipőcz, Robitaille, Tollerud,
		Vinícius, Deil, Barbary, Wilson, Busko, Günther, Cara, Conseil, Bostroem,
		Droettboom, Bray, Bratholm, Lim, Barentsen, Craig, Pascual, Perren, Greco,
		Donath, de~Val-Borro, Kerzendorf, Bach, Weaver, D'Eugenio, Souchereau, \&
		Ferreira}]{Bradley2020}
	Bradley, L., Sipőcz, B., Robitaille, T., {et~al.} 2020, astropy/photutils:
	1.0.1
	
	\bibitem[{Buta {et~al.}(2015)Buta, Sheth, Athanassoula, Bosma, Knapen,
		Laurikainen, Salo, Elmegreen, Ho, Zaritsky, Courtois, Hinz, Mu{\~n}oz-Mateos,
		Kim, Regan, Gadotti, Gil~de Paz, Laine, Men{\'e}ndez-Delmestre, Comer{\'o}n,
		Erroz~Ferrer, Seibert, Mizusawa, Holwerda, \& Madore}]{Buta2015}
	Buta, R.~J., Sheth, K., Athanassoula, E., {et~al.} 2015, \apjs, 217, 32
	
	\bibitem[{Carraro {et~al.}(2015)Carraro, V{\'a}zquez, Costa, Ahumada, \&
		Giorgi}]{Carraro2015a}
	Carraro, G., V{\'a}zquez, R.~A., Costa, E., Ahumada, J.~A., \& Giorgi, E.~E.
	2015, \aj, 149, 12
	
	\bibitem[{Chamba {et~al.}(2022)Chamba, Trujillo, \& Knapen}]{Chamba2022}
	Chamba, N., Trujillo, I., \& Knapen, J.~H. 2022, \aap, 667, A87
	
	\bibitem[{Chrob{\'a}kov{\'a} {et~al.}(2022)Chrob{\'a}kov{\'a}, Nagy, \&
		L{\'o}pez-Corredoira}]{Chrobakova2022}
	Chrob{\'a}kov{\'a}, {\v{Z}}., Nagy, R., \& L{\'o}pez-Corredoira, M. 2022, \aap,
	664, A58
	
	\bibitem[{Comer{\'{o}}n {et~al.}(2012)Comer{\'{o}}n, Elmegreen, Salo,
		Laurikainen, Athanassoula, Bosma, Knapen, Gadotti, Sheth, Hinz, Regan, Gil~de
		Paz, Mu{\~n}oz-Mateos, Men{\'e}ndez-Delmestre, Seibert, Kim, Mizusawa, Laine,
		Ho, \& Holwerda}]{Comeron2012}
	Comer{\'{o}}n, S., Elmegreen, B.~G., Salo, H., {et~al.} 2012, \apj, 759, 98
	
	\bibitem[{Comer{\'{o}}n {et~al.}(2018)Comer{\'{o}}n, Salo, \&
		Knapen}]{Comeron2017}
	Comer{\'{o}}n, S., Salo, H., \& Knapen, J.~H. 2018, \aap, 610, A5
	
	\bibitem[{Das {et~al.}(2020)Das, Sardone, Leroy, Mathur, Gallagher, Pingel,
		Pisano, \& Heald}]{Das2020}
	Das, S., Sardone, A., Leroy, A.~K., {et~al.} 2020, \apj, 898, 15
	
	\bibitem[{de~Jong(2008)}]{deJong2008}
	de~Jong, R.~S. 2008, \mnras, 388, 1521
	
	\bibitem[{de~Jong {et~al.}(2007)de~Jong, Seth, Radburn-Smith, Bell, Brown,
		Bullock, Courteau, Dalcanton, Ferguson, Goudfrooij, Holfeltz, Holwerda,
		Purcell, Sick, \& Zucker}]{deJong2007}
	de~Jong, R.~S., Seth, A.~C., Radburn-Smith, D.~J., {et~al.} 2007, \apjl, 667,
	L49
	
	\bibitem[{Debattista {et~al.}(2017)Debattista, Ro{\v{s}}kar, \&
		Loebman}]{Debattista2017}
	Debattista, V.~P., Ro{\v{s}}kar, R., \& Loebman, S.~R. 2017, The Impact of
	Stellar Migration on Disk Outskirts, ed. G.~d.~P. Knapen, Lee (Astrophysics
	and Space Science Library, Springer International Publishing, Cham,
	Switzerland), 77--114
	
	\bibitem[{D{\'\i}az-Garc{\'\i}a {et~al.}(2022)D{\'\i}az-Garc{\'\i}a,
		Comer{\'o}n, Courteau, Watkins, Knapen, \& Rom{\'a}n}]{DiazGarcia2022}
	D{\'\i}az-Garc{\'\i}a, S., Comer{\'o}n, S., Courteau, S., {et~al.} 2022, \aap,
	667, A109
	
	\bibitem[{Ding {et~al.}(2021)Ding, Xue, Yang, Zhao, Zhang, \& Zhu}]{Ding2021}
	Ding, P.-J., Xue, X.-X., Yang, C., {et~al.} 2021, \aj, 162, 112
	
	\bibitem[{Duc {et~al.}(2015)Duc, Cuillandre, Karabal, Cappellari, Alatalo,
		Blitz, Bournaud, Bureau, Crocker, Davies, Davis, de~Zeeuw, Emsellem,
		Khochfar, Krajnovi{\'{c}}, Kuntschner, McDermid, Michel-Dansac, Morganti,
		Naab, Oosterloo, Paudel, Sarzi, Scott, Serra, Weijmans, \& Young}]{Duc2015}
	Duc, P.-A., Cuillandre, J.-C., Karabal, E., {et~al.} 2015, \mnras, 446, 120
	
	\bibitem[{Erben {et~al.}(2005)Erben, Schirmer, Dietrich, Cordes, Haberzettl,
		Hetterscheidt, Hildebrandt, Schmithuesen, Schneider, Simon, Deul, Hook,
		Kaiser, Radovich, Benoist, Nonino, Olsen, Prandoni, Wichmann, Zaggia, Bomans,
		Dettmar, \& Miralles}]{Erben2005}
	Erben, T., Schirmer, M., Dietrich, J.~P., {et~al.} 2005, Astronomische
	Nachrichten, 326, 432
	
	\bibitem[{Erwin(2015)}]{Erwin2015}
	Erwin, P. 2015, \apj, 799, 226
	
	\bibitem[{Feast {et~al.}(2014)Feast, Menzies, Matsunaga, \&
		Whitelock}]{Feast2014}
	Feast, M.~W., Menzies, J.~W., Matsunaga, N., \& Whitelock, P.~A. 2014, \nat,
	509, 342
	
	\bibitem[{{Gaia Collaboration} {et~al.}(2021){Gaia Collaboration}, Brown,
		Vallenari, Prusti, de~Bruijne, Babusiaux, Biermann, Creevey, Evans, Eyer,
		Hutton, Jansen, Jordi, Klioner, Lammers, Lindegren, Luri, Mignard, Panem,
		Pourbaix, Randich, Sartoretti, Soubiran, Walton, Arenou, Bailer-Jones,
		Bastian, Cropper, Drimmel, Katz, Lattanzi, van Leeuwen, Bakker, Cacciari,
		Casta{\~n}eda, De~Angeli, Ducourant, Fabricius, Fouesneau, Fr{\'e}mat,
		Guerra, Guerrier, Guiraud, Jean-Antoine~Piccolo, Masana, Messineo, Mowlavi,
		Nicolas, Nienartowicz, Pailler, Panuzzo, Riclet, Roux, Seabroke, Sordo,
		Tanga, Th{\'e}venin, Gracia-Abril, Portell, Teyssier, Altmann, Andrae,
		Bellas-Velidis, Benson, Berthier, Blomme, Brugaletta, Burgess, Busso, Carry,
		Cellino, Cheek, Clementini, Damerdji, Davidson, Delchambre, Dell'Oro,
		Fern{\'a}ndez-Hern{\'a}ndez, Galluccio, Garc{\'\i}a-Lario, Garcia-Reinaldos,
		Gonz{\'a}lez-N{\'u}{\~n}ez, Gosset, Haigron, Halbwachs, Hambly, Harrison,
		Hatzidimitriou, Heiter, Hern{\'a}ndez, Hestroffer, Hodgkin, Holl, Jan{\ss}en,
		Jevardat~de Fombelle, Jordan, Krone-Martins, Lanzafame, L{\"o}ffler, Lorca,
		Manteiga, Marchal, Marrese, Moitinho, Mora, Muinonen, Osborne, Pancino,
		Pauwels, Petit, Recio-Blanco, Richards, Riello, Rimoldini, Robin, Roegiers,
		Rybizki, Sarro, Siopis, Smith, Sozzetti, Ulla, Utrilla, van Leeuwen, van
		Reeven, Abbas, Abreu~Aramburu, Accart, Aerts, Aguado, Ajaj, Altavilla,
		{\'A}lvarez, {\'A}lvarez Cid-Fuentes, Alves, Anderson, Anglada~Varela,
		Antoja, Audard, Baines, Baker, Balaguer-N{\'u}{\~n}ez, Balbinot, Balog,
		Barache, Barbato, Barros, Barstow, Bartolom{\'e}, Bassilana, Bauchet,
		Baudesson-Stella, Becciani, Bellazzini, Bernet, Bertone, Bianchi,
		Blanco-Cuaresma, Boch, Bombrun, Bossini, Bouquillon, Bragaglia, Bramante,
		Breedt, Bressan, Brouillet, Bucciarelli, Burlacu, Busonero, Butkevich, Buzzi,
		Caffau, Cancelliere, C{\'a}novas, Cantat-Gaudin, Carballo, Carlucci,
		Carnerero, Carrasco, Casamiquela, Castellani, Castro-Ginard, Castro~Sampol,
		Chaoul, Charlot, Chemin, Chiavassa, Cioni, Comoretto, Cooper, Cornez, Cowell,
		Crifo, Crosta, Crowley, Dafonte, Dapergolas, David, David, de~Laverny,
		De~Luise, De~March, De~Ridder, de~Souza, de~Teodoro, de~Torres, del Peloso,
		del Pozo, Delbo, Delgado, Delgado, Delisle, Di~Matteo, Diakite, Diener,
		Distefano, Dolding, Eappachen, Edvardsson, Enke, Esquej, Fabre, Fabrizio,
		Faigler, Fedorets, Fernique, Fienga, Figueras, Fouron, Fragkoudi, Fraile,
		Franke, Gai, Garabato, Garcia-Gutierrez, Garc{\'\i}a-Torres, Garofalo,
		Gavras, Gerlach, Geyer, Giacobbe, Gilmore, Girona, Giuffrida, Gomel, Gomez,
		Gonzalez-Santamaria, Gonz{\'a}lez-Vidal, Granvik, Guti{\'e}rrez-S{\'a}nchez,
		Guy, Hauser, Haywood, Helmi, Hidalgo, Hilger, H{\l}adczuk, Hobbs, Holland,
		Huckle, Jasniewicz, Jonker, Juaristi~Campillo, Julbe, Karbevska, Kervella,
		Khanna, Kochoska, Kontizas, Kordopatis, Korn, Kostrzewa-Rutkowska,
		Kruszy{\'n}ska, Lambert, Lanza, Lasne, Le~Campion, Le~Fustec, Lebreton,
		Lebzelter, Leccia, Leclerc, Lecoeur-Taibi, Liao, Licata, Lindstr{\o}m,
		Lister, Livanou, Lobel, Madrero~Pardo, Managau, Mann, Marchant, Marconi,
		Marcos~Santos, Marinoni, Marocco, Marshall, Martin~Polo, Mart{\'\i}n-Fleitas,
		Masip, Massari, Mastrobuono-Battisti, Mazeh, McMillan, Messina, Michalik,
		Millar, Mints, Molina, Molinaro, Moln{\'a}r, Montegriffo, Mor, Morbidelli,
		Morel, Morris, Mulone, Munoz, Muraveva, Murphy, Musella, Noval,
		Ord{\'e}novic, Orr{\`u}, Osinde, Pagani, Pagano, Palaversa, Palicio, Panahi,
		Pawlak, Pe{\~n}alosa~Esteller, Penttil{\"a}, Piersimoni, Pineau, Plachy,
		Plum, Poggio, Poretti, Poujoulet, Pr{\v{s}}a, Pulone, Racero, Ragaini,
		Rainer, Raiteri, Rambaux, Ramos, Ramos-Lerate, Re~Fiorentin, Regibo,
		Reyl{\'e}, Ripepi, Riva, Rixon, Robichon, Robin, Roelens, Rohrbasser,
		Romero-G{\'o}mez, Rowell, Royer, Rybicki, Sadowski, Sagrist{\`a Sell{\'e}s},
		Sahlmann, Salgado, Salguero, Samaras, Sanchez~Gimenez, Sanna, Santove{\~n}a,
		Sarasso, Schultheis, Sciacca, Segol, Segovia, S{\'e}gransan, Semeux, Shahaf,
		Siddiqui, Siebert, Siltala, Slezak, Smart, Solano, Solitro, Souami, Souchay,
		Spagna, Spoto, Steele, Steidelm{\"u}ller, Stephenson, S{\"u}veges, Szabados,
		Szegedi-Elek, Taris, Tauran, Taylor, Teixeira, Thuillot, Tonello, Torra,
		Torra, Turon, Unger, Vaillant, van Dillen, Vanel, Vecchiato, Viala, Vicente,
		Voutsinas, Weiler, Wevers, Wyrzykowski, Yoldas, Yvard, Zhao, Zorec, Zucker,
		Zurbach, \& Zwitter}]{Collaboration2021}
	{Gaia Collaboration}, Brown, A. G.~A., Vallenari, A., {et~al.} 2021, \aap, 649,
	A1
	
	\bibitem[{Gilhuly {et~al.}(2020)Gilhuly, Hendel, Merritt, Abraham, Danieli,
		Lokhorst, Liu, van Dokkum, Conroy, \& Greco}]{Gilhuly2020}
	Gilhuly, C., Hendel, D., Merritt, A., {et~al.} 2020, \apj, 897, 108
	
	\bibitem[{Gregory \& Thompson(1977)}]{Gregory1977}
	Gregory, S.~A. \& Thompson, L.~A. 1977, \apj, 213, 345
	
	\bibitem[{Hao {et~al.}(2011)Hao, Kennicutt, Johnson, Calzetti, Dale, \&
		Moustakas}]{Hao2011}
	Hao, C.-N., Kennicutt, R.~C., Johnson, B.~D., {et~al.} 2011, \apj, 741, 124
	
	\bibitem[{Heald {et~al.}(2012)Heald, J{\'o}zsa, Serra, Zschaechner, Rand,
		Fraternali, Oosterloo, Walterbos, J{\"u}tte, \& Gentile}]{Heald2012}
	Heald, G., J{\'o}zsa, G., Serra, P., {et~al.} 2012, \aap, 544, C1
	
	\bibitem[{Helmboldt {et~al.}(2004)Helmboldt, Walterbos, Bothun, O'Neil, \&
		de~Blok}]{Helmboldt2004}
	Helmboldt, J.~F., Walterbos, R. A.~M., Bothun, G.~D., O'Neil, K., \& de~Blok,
	W. J.~G. 2004, \apj, 613, 914
	
	\bibitem[{Hunter {et~al.}(2012)Hunter, Ficut-Vicas, Ashley, Brinks, Cigan,
		Elmegreen, Heesen, Herrmann, Johnson, Oh, Rupen, Schruba, Simpson, Walter,
		Westpfahl, Young, \& Zhang}]{Hunter2012}
	Hunter, D.~A., Ficut-Vicas, D., Ashley, T., {et~al.} 2012, \aj, 144, 134
	
	\bibitem[{Infante-Sainz {et~al.}(2020)Infante-Sainz, Trujillo, \&
		Rom{\'a}n}]{InfanteSainz2020}
	Infante-Sainz, R., Trujillo, I., \& Rom{\'a}n, J. 2020, \mnras, 491, 5317
	
	\bibitem[{Iodice {et~al.}(2019{\natexlab{a}})Iodice, Sarzi, Bittner, Coccato,
		Costantin, Corsini, van~de Ven, de~Zeeuw, Falc{\'o}n-Barroso, Gadotti,
		Lyubenova, Mart{\'\i}n-Navarro, McDermid, Nedelchev, Pinna, Pizzella,
		Spavone, \& Viaene}]{Iodice2019}
	Iodice, E., Sarzi, M., Bittner, A., {et~al.} 2019{\natexlab{a}}, \aap, 627,
	A136
	
	\bibitem[{Iodice {et~al.}(2019{\natexlab{b}})Iodice, Spavone, Capaccioli,
		Peletier, van~de Ven, Napolitano, Hilker, Mieske, Smith, Pasquali, Limatola,
		Grado, Venhola, Cantiello, Paolillo, Falcon-Barroso, D'Abrusco, \&
		Schipani}]{Iodice2019b}
	Iodice, E., Spavone, M., Capaccioli, M., {et~al.} 2019{\natexlab{b}}, \aap,
	623, A1
	
	\bibitem[{James {et~al.}(2005)James, Shane, Knapen, Etherton, \&
		Percival}]{James2005}
	James, P.~A., Shane, N.~S., Knapen, J.~H., Etherton, J., \& Percival, S.~M.
	2005, \aap, 429, 851
	
	\bibitem[{Joye \& Mandel(2003)}]{Joye2003}
	Joye, W.~A. \& Mandel, E. 2003, in Astronomical Society of the Pacific
	Conference Series, Vol. 295, Astronomical Data Analysis Software and Systems
	XII, ed. H.~E. {Payne}, R.~I. {Jedrzejewski}, \& R.~N. {Hook}, 489
	
	\bibitem[{Kado-Fong {et~al.}(2018)Kado-Fong, Greene, Hendel, Price-Whelan,
		Greco, Goulding, Huang, Johnston, Komiyama, Lee, Lust, Strauss, \&
		Tanaka}]{KadoFong2018}
	Kado-Fong, E., Greene, J.~E., Hendel, D., {et~al.} 2018, \apj, 866, 103
	
	\bibitem[{Kalberla {et~al.}(2014)Kalberla, Kerp, Dedes, \& Haud}]{Kalberla2014}
	Kalberla, P. M.~W., Kerp, J., Dedes, L., \& Haud, U. 2014, \apj, 794, 90
	
	\bibitem[{Kasparova {et~al.}(2020)Kasparova, Katkov, \&
		Chilingarian}]{Kasparova2020}
	Kasparova, A.~V., Katkov, I.~Y., \& Chilingarian, I.~V. 2020, \mnras, 493, 5464
	
	\bibitem[{Kaviraj {et~al.}(2017)Kaviraj, Laigle, Kimm, Devriendt, Dubois,
		Pichon, Slyz, Chisari, \& Peirani}]{Kaviraj2017}
	Kaviraj, S., Laigle, C., Kimm, T., {et~al.} 2017, \mnras, 467, 4739
	
	\bibitem[{Kazantzidis {et~al.}(2008)Kazantzidis, Bullock, Zentner, Kravtsov, \&
		Moustakas}]{Kazantzidis2008}
	Kazantzidis, S., Bullock, J.~S., Zentner, A.~R., Kravtsov, A.~V., \& Moustakas,
	L.~A. 2008, \apj, 688, 254
	
	\bibitem[{Kennicutt \& Kent(1983)}]{Kennicutt1983}
	Kennicutt, R.~C., J. \& Kent, S.~M. 1983, \aj, 88, 1094
	
	\bibitem[{Kennicutt(1998)}]{Kennicutt1998}
	Kennicutt, Robert~C., J. 1998, \araa, 36, 189
	
	\bibitem[{Kennicutt {et~al.}(2009)Kennicutt, Hao, Calzetti, Moustakas, Dale,
		Bendo, Engelbracht, Johnson, \& Lee}]{Kennicutt2009}
	Kennicutt, Robert~C., J., Hao, C.-N., Calzetti, D., {et~al.} 2009, \apj, 703,
	1672
	
	\bibitem[{Kennicutt {et~al.}(2008)Kennicutt, Lee, Funes, S., Sakai, \&
		Akiyama}]{Kennicutt2008}
	Kennicutt, Robert~C., J., Lee, J.~C., Funes, J.~G., {et~al.} 2008, \apjs, 178,
	247
	
	\bibitem[{Kennicutt(1989)}]{Kennicutt1989}
	Kennicutt, Jr., R.~C. 1989, \apj, 344, 685
	
	\bibitem[{Kim {et~al.}(2012)Kim, Sheth, Hinz, Lee, Zaritsky, Gadotti, Knapen,
		Schinnerer, Ho, Laurikainen, Salo, Athanassoula, Bosma, de~Swardt,
		Mu{\~n}oz-Mateos, Madore, Comer{\'o}n, Regan, Men{\'e}ndez-Delmestre, Gil~de
		Paz, Seibert, Laine, Erroz-Ferrer, \& Mizusawa}]{Kim2012}
	Kim, T., Sheth, K., Hinz, J.~L., {et~al.} 2012, \apj, 753, 43
	
	\bibitem[{{Knapen} {et~al.}(2004){Knapen}, {Stedman}, {Bramich}, {Folkes}, \&
		{Bradley}}]{Knapen2004}
	{Knapen}, J.~H., {Stedman}, S., {Bramich}, D.~M., {Folkes}, S.~L., \&
	{Bradley}, T.~R. 2004, \aap, 426, 1135
	
	\bibitem[{Kormendy \& Bender(2019)}]{Kormendy2019}
	Kormendy, J. \& Bender, R. 2019, \apj, 872, 106
	
	\bibitem[{{Kormendy} \& {Kennicutt}(2004)}]{Kormendy2004}
	{Kormendy}, J. \& {Kennicutt}, Jr., R.~C. 2004, \araa, 42, 603
	
	\bibitem[{Kroupa \& Weidner(2003)}]{Kroupa2003}
	Kroupa, P. \& Weidner, C. 2003, \apj, 598, 1076
	
	\bibitem[{Lang {et~al.}(2010)Lang, Hogg, Mierle, Blanton, \& Roweis}]{Lang2010}
	Lang, D., Hogg, D.~W., Mierle, K., Blanton, M., \& Roweis, S. 2010, \aj, 139,
	1782
	
	\bibitem[{Makarov {et~al.}(2014)Makarov, Prugniel, Terekhova, Courtois, \&
		Vauglin}]{Makarov2014}
	Makarov, D., Prugniel, P., Terekhova, N., Courtois, H., \& Vauglin, I. 2014,
	\aap, 570, A13
	
	\bibitem[{Martin {et~al.}(2005)Martin, Fanson, Schiminovich, Morrissey,
		Friedman, Barlow, Conrow, Grange, Jelinsky, Milliard, Siegmund, Bianchi,
		Byun, Donas, Forster, Heckman, Lee, Madore, Malina, Neff, Rich, Small,
		Surber, Szalay, Welsh, \& Wyder}]{Martin2005}
	Martin, D.~C., Fanson, J., Schiminovich, D., {et~al.} 2005, \apjl, 619, L1
	
	\bibitem[{Mart{\'{\i}}n-Navarro {et~al.}(2012)Mart{\'{\i}}n-Navarro, Bakos,
		Trujillo, Knapen, Athanassoula, Bosma, Comer{\'o}n, Elmegreen, Erroz-Ferrer,
		Gadotti, Gil~de Paz, Hinz, Ho, Holwerda, Kim, Laine, Laurikainen,
		Men{\'e}ndez-Delmestre, Mizusawa, Mu{\~n}oz-Mateos, Regan, Salo, Seibert, \&
		Sheth}]{Martin-Navarro2012}
	Mart{\'{\i}}n-Navarro, I., Bakos, J., Trujillo, I., {et~al.} 2012, \mnras, 427,
	1102
	
	\bibitem[{Mart{\'{\i}}nez-Delgado {et~al.}(2009)Mart{\'{\i}}nez-Delgado,
		Pohlen, Gabany, Majewski, Pe{\~n}arrubia, \& Palma}]{Martinez-Delgado2009}
	Mart{\'{\i}}nez-Delgado, D., Pohlen, M., Gabany, R.~J., {et~al.} 2009, \apj,
	692, 955
	
	\bibitem[{Mart{\'\i}nez-Lombilla {et~al.}(2023)Mart{\'\i}nez-Lombilla, Brough,
		Montes, Baena-Gall{\'e}, Akhlaghi, Infante-Sainz, Driver, Holwerda, Pimbblet,
		\& Robotham}]{MartinezLombilla2023}
	Mart{\'\i}nez-Lombilla, C., Brough, S., Montes, M., {et~al.} 2023, \mnras, 518,
	1195
	
	\bibitem[{Mart{\'\i}nez-Lombilla \& Knapen(2019)}]{MartinezLombilla2019b}
	Mart{\'\i}nez-Lombilla, C. \& Knapen, J.~H. 2019, \aap, 629, A12
	
	\bibitem[{{Mart{\'\i}nez-Lombilla} {et~al.}(2019){Mart{\'\i}nez-Lombilla},
		{Trujillo}, \& {Knapen}}]{MartinezLombilla2019a}
	{Mart{\'\i}nez-Lombilla}, C., {Trujillo}, I., \& {Knapen}, J.~H. 2019, \mnras,
	483, 664
	
	\bibitem[{McMullin {et~al.}(2007)McMullin, Waters, Schiebel, Young, \&
		Golap}]{McMullin2007}
	McMullin, J.~P., Waters, B., Schiebel, D., Young, W., \& Golap, K. 2007, in
	Astronomical Society of the Pacific Conference Series, Vol. 376, Astronomical
	Data Analysis Software and Systems XVI, ed. R.~A. {Shaw}, F.~{Hill}, \& D.~J.
	{Bell}, 127
	
	\bibitem[{Michard(2002)}]{Michard2002}
	Michard, R. 2002, \aap, 384, 763
	
	\bibitem[{Mihos {et~al.}(2017)Mihos, Harding, Feldmeier, Rudick, Janowiecki,
		Morrison, Slater, \& Watkins}]{Mihos2017}
	Mihos, J.~C., Harding, P., Feldmeier, J.~J., {et~al.} 2017, \apj, 834, 16
	
	\bibitem[{Minchev {et~al.}(2015)Minchev, Martig, Streich, Scannapieco, de~Jong,
		\& Steinmetz}]{Minchev2015}
	Minchev, I., Martig, M., Streich, D., {et~al.} 2015, \apjl, 804, L9
	
	\bibitem[{Montes(2022)}]{Montes2022}
	Montes, M. 2022, Nature Astronomy, 6, 308
	
	\bibitem[{Montes {et~al.}(2020)Montes, Infante-Sainz, Madrigal-Aguado,
		Rom{\'a}n, Monelli, Borlaff, \& Trujillo}]{Montes2020}
	Montes, M., Infante-Sainz, R., Madrigal-Aguado, A., {et~al.} 2020, \apj, 904,
	114
	
	\bibitem[{Montes \& Trujillo(2018)}]{Montes2018}
	Montes, M. \& Trujillo, I. 2018, \mnras, 474, 917
	
	\bibitem[{Montes \& Trujillo(2019)}]{Montes2019}
	Montes, M. \& Trujillo, I. 2019, \mnras, 482, 2838
	
	\bibitem[{Morrissey {et~al.}(2007)Morrissey, Conrow, Barlow, Small, Seibert,
		Wyder, Budav{\'a}ri, Arnouts, Friedman, Forster, Martin, Neff, Schiminovich,
		Bianchi, Donas, Heckman, Lee, Madore, Milliard, Rich, Szalay, Welsh, \&
		Yi}]{Morrissey2007}
	Morrissey, P., Conrow, T., Barlow, T.~A., {et~al.} 2007, \apjs, 173, 682
	
	\bibitem[{Mosenkov {et~al.}(2020)Mosenkov, Rich, Koch, Brosch, Thilker,
		Rom{\'a}n, M{\"u}ller, Smirnov, \& Usachev}]{Mosenkov2020}
	Mosenkov, A., Rich, R.~M., Koch, A., {et~al.} 2020, \mnras, 494, 1751
	
	\bibitem[{Naeslund \& Joersaeter(1997)}]{Naeslund1997}
	Naeslund, M. \& Joersaeter, S. 1997, \aap, 325, 915
	
	\bibitem[{Niklas {et~al.}(1997)Niklas, Klein, \& Wielebinski}]{Niklas1997}
	Niklas, S., Klein, U., \& Wielebinski, R. 1997, \aap, 322, 19
	
	\bibitem[{Oliphant(2006)}]{Oliphant2006}
	Oliphant, T.~E. 2006, All Faculty Publications, 278
	
	\bibitem[{Oosterloo {et~al.}(2007)Oosterloo, Fraternali, \&
		Sancisi}]{Oosterloo2007}
	Oosterloo, T., Fraternali, F., \& Sancisi, R. 2007, \aj, 134, 1019
	
	\bibitem[{Padilla {et~al.}(2019)Padilla, Castander, Alarc{\'o}n, Aleksic,
		Ballester, Cabayol, Cardiel-Sas, Carretero, Casas, Castilla, Crocce, Delfino,
		D{\'\i}az, Eriksen, Fern{\'a}ndez, Fosalba, Garc{\'\i}a-Bellido,
		Gazta{\~n}aga, Gaweda, Gra{\~n}ena, Mar{\'\i}a~{\'I}lla, Jim{\'e}nez,
		L{\'o}pez, Mart{\'\i}, Miquel, Neissner, P{\'\i}o, S{\'a}nchez, Serrano,
		Sevilla-Noarbe, Tallada, Tonello, \& de~Vicente}]{Padilla2019}
	Padilla, C., Castander, F.~J., Alarc{\'o}n, A., {et~al.} 2019, \aj, 157, 246
	
	\bibitem[{Peters {et~al.}(2017)Peters, van~der Kruit, Knapen, Trujillo, Fliri,
		Cisternas, \& Kelvin}]{Peters2017}
	Peters, S. P.~C., van~der Kruit, P.~C., Knapen, J.~H., {et~al.} 2017, \mnras,
	470, 427
	
	\bibitem[{Pinna {et~al.}(2019{\natexlab{a}})Pinna, Falc{\'o}n-Barroso, Martig,
		Coccato, Corsini, de~Zeeuw, Gadotti, Iodice, Leaman, Lyubenova,
		Mart{\'\i}n-Navarro, Morelli, Sarzi, van~de Ven, Viaene, \&
		McDermid}]{Pinna2019b}
	Pinna, F., Falc{\'o}n-Barroso, J., Martig, M., {et~al.} 2019{\natexlab{a}},
	\aap, 625, A95
	
	\bibitem[{Pinna {et~al.}(2019{\natexlab{b}})Pinna, Falc{\'o}n-Barroso, Martig,
		Sarzi, Coccato, Iodice, Corsini, de~Zeeuw, Gadotti, Leaman, Lyubenova,
		McDermid, Minchev, Morelli, van~de Ven, \& Viaene}]{Pinna2019a}
	Pinna, F., Falc{\'o}n-Barroso, J., Martig, M., {et~al.} 2019{\natexlab{b}},
	\aap, 623, A19
	
	\bibitem[{{Pohlen} \& {Trujillo}(2006)}]{PohlenTrujillo2006}
	{Pohlen}, M. \& {Trujillo}, I. 2006, \aap, 454, 759
	
	\bibitem[{Qu {et~al.}(2011)Qu, Matteo, Lehnert, \& van Driel}]{Qu2011}
	Qu, Y., Matteo, P.~D., Lehnert, M.~D., \& van Driel, W. 2011, \aap, 530, A10
	
	\bibitem[{Radburn-Smith {et~al.}(2014)Radburn-Smith, de~Jong, Streich, Bell,
		Dalcanton, Dolphin, Stilp, Monachesi, Holwerda, \&
		Bailin}]{Radburn-Smith2014}
	Radburn-Smith, D.~J., de~Jong, R.~S., Streich, D., {et~al.} 2014, \apj, 780,
	105
	
	\bibitem[{Radburn-Smith {et~al.}(2012)Radburn-Smith, Ro{\v{s}}kar, Debattista,
		Dalcanton, Streich, de~Jong, Vlaji{\'c}, Holwerda, Purcell, Dolphin, \&
		Zucker}]{RadburnSmith2012}
	Radburn-Smith, D.~J., Ro{\v{s}}kar, R., Debattista, V.~P., {et~al.} 2012, \apj,
	753, 138
	
	\bibitem[{Rix {et~al.}(2004)Rix, Barden, Beckwith, Bell, Borch, Caldwell,
		H{\"a}ussler, Jahnke, Jogee, McIntosh, Meisenheimer, Peng, Sanchez,
		Somerville, Wisotzki, \& Wolf}]{Rix2004}
	Rix, H.-W., Barden, M., Beckwith, S. V.~W., {et~al.} 2004, \apjs, 152, 163
	
	\bibitem[{Rom{\'a}n {et~al.}(2020)Rom{\'a}n, Trujillo, \& Montes}]{Roman2020}
	Rom{\'a}n, J., Trujillo, I., \& Montes, M. 2020, \aap, 644, A42
	
	\bibitem[{Ro{\v{s}}kar {et~al.}(2010)Ro{\v{s}}kar, Debattista, Brooks, Quinn,
		Brook, Governato, Dalcanton, \& Wadsley}]{Roskar2010}
	Ro{\v{s}}kar, R., Debattista, V.~P., Brooks, A.~M., {et~al.} 2010, \mnras, 408,
	783
	
	\bibitem[{Ro{\v s}kar {et~al.}(2008{\natexlab{a}})Ro{\v s}kar, Debattista,
		Quinn, Stinson, \& Wadsley}]{Rovskar2008b}
	Ro{\v s}kar, R., Debattista, V.~P., Quinn, T.~R., Stinson, G.~S., \& Wadsley,
	J. 2008{\natexlab{a}}, \apjl, 684, L79
	
	\bibitem[{Ro{\v s}kar {et~al.}(2008{\natexlab{b}})Ro{\v s}kar, Debattista,
		Stinson, Quinn, Kaufmann, \& Wadsley}]{Rovskar2008a}
	Ro{\v s}kar, R., Debattista, V.~P., Stinson, G.~S., {et~al.}
	2008{\natexlab{b}}, \apjl, 675, L65
	
	\bibitem[{Rupen(1991)}]{Rupen1991}
	Rupen, M.~P. 1991, \aj, 102, 48
	
	\bibitem[{Saifollahi {et~al.}(2022)Saifollahi, Zaritsky, Trujillo, Peletier,
		Knapen, Amorisco, Beasley, \& Donnerstein}]{Saifollahi2022}
	Saifollahi, T., Zaritsky, D., Trujillo, I., {et~al.} 2022, \mnras, 511, 4633
	
	\bibitem[{{S{\'a}nchez-Gallego} {et~al.}(2012){S{\'a}nchez-Gallego}, {Knapen},
		{Wilson}, {Barmby}, {Azimlu}, \& {Courteau}}]{Sanchez-Gallego2012}
	{S{\'a}nchez-Gallego}, J.~R., {Knapen}, J.~H., {Wilson}, C.~D., {et~al.} 2012,
	\mnras, 422, 3208
	
	\bibitem[{Sandin(2014)}]{Sandin2014}
	Sandin, C. 2014, \aap, 567, A97
	
	\bibitem[{Sandin(2015)}]{Sandin2015}
	Sandin, C. 2015, \aap, 567, A106
	
	\bibitem[{Scannapieco {et~al.}(2009)Scannapieco, White, Springel, \&
		Tissera}]{Scannapieco2009}
	Scannapieco, C., White, S. D.~M., Springel, V., \& Tissera, P.~B. 2009, \mnras,
	396, 696
	
	\bibitem[{Schaye(2004)}]{Schaye2004}
	Schaye, J. 2004, \apj, 609, 667
	
	\bibitem[{Schirmer(2013)}]{Schirmer2013}
	Schirmer, M. 2013, \apjs, 209, 21
	
	\bibitem[{Schlafly \& Finkbeiner(2011)}]{Schlafly2011}
	Schlafly, E.~F. \& Finkbeiner, D.~P. 2011, \apj, 737, 103
	
	\bibitem[{Sellwood(2014)}]{Sellwood2014}
	Sellwood, J.~A. 2014, Reviews of Modern Physics, 86, 1
	
	\bibitem[{{Sersic}(1968)}]{Sersic1968}
	{Sersic}, J.~L. 1968, 
	Nature), 342--354
	
	\bibitem[{Slater {et~al.}(2009)Slater, Harding, \& Mihos}]{Slater2009}
	Slater, C.~T., Harding, P., \& Mihos, J.~C. 2009, \pasp, 121, 1267
	
	\bibitem[{Trujillo {et~al.}(2020)Trujillo, Chamba, \& Knapen}]{Trujillo2020}
	Trujillo, I., Chamba, N., \& Knapen, J.~H. 2020, \mnras, 493, 87
	
	\bibitem[{Trujillo {et~al.}(2021)Trujillo, D'Onofrio, Zaritsky,
		Madrigal-Aguado, Chamba, Golini, Akhlaghi, Sharbaf, Infante-Sainz, Rom{\'a}n,
		Morales-Socorro, Sand, \& Martin}]{Trujillo2021}
	Trujillo, I., D'Onofrio, M., Zaritsky, D., {et~al.} 2021, \aap, 654, A40
	
	\bibitem[{Trujillo \& Fliri(2016)}]{TrujilloFliri2016}
	Trujillo, I. \& Fliri, J. 2016, \apj, 823, 123
	
	\bibitem[{{van der Kruit}(1979)}]{vanderKruit1979}
	{van der Kruit}, P.~C. 1979, \aaps, 38, 15
	
	\bibitem[{{van der Kruit}(1987)}]{vanderkruit1987a}
	{van der Kruit}, P.~C. 1987, \aap, 173, 59
	
	\bibitem[{van~der Kruit \& Freeman(2011)}]{KruitFreeman2011}
	van~der Kruit, P.~C. \& Freeman, K.~C. 2011, \araa, 49, 301
	
	\bibitem[{{van der Kruit} \& {Searle}(1981{\natexlab{a}})}]{vanderKruit1981a}
	{van der Kruit}, P.~C. \& {Searle}, L. 1981{\natexlab{a}}, \aap, 95, 105
	
	\bibitem[{{van der Kruit} \& {Searle}(1981{\natexlab{b}})}]{vanderKruit1981b}
	{van der Kruit}, P.~C. \& {Searle}, L. 1981{\natexlab{b}}, \aap, 95, 116
	
	\bibitem[{Villalobos \& Helmi(2008)}]{Villalobos2008}
	Villalobos, {\'A}. \& Helmi, A. 2008, \mnras, 391, 1806
	
	\bibitem[{Wiegert {et~al.}(2015)Wiegert, Irwin, Miskolczi, Schmidt, Mora,
		Damas-Segovia, Stein, English, Rand, Santistevan, Walterbos, Krause, Beck,
		Dettmar, Kepley, Wezgowiec, Wang, Heald, Li, MacGregor, Johnson, Strong,
		DeSouza, \& Porter}]{Wiegert2015}
	Wiegert, T., Irwin, J., Miskolczi, A., {et~al.} 2015, \aj, 150, 81
	
	\bibitem[{Wilson {et~al.}(2009)Wilson, Warren, Israel, Serjeant, Bendo, Brinks,
		Clements, Courteau, Irwin, Knapen, Leech, Matthews, M{\"u}hle, Mortier,
		Petitpas, Sinukoff, Spekkens, Tan, Tilanus, Usero, van~der Werf, Wiegert, \&
		Zhu}]{Wilson2009}
	Wilson, C.~D., Warren, B.~E., Israel, F.~P., {et~al.} 2009, \apj, 693, 1736
	
	\bibitem[{Wu {et~al.}(2002)Wu, Burstein, Deng, Zhou, Shang, Zheng, Chen, Su,
		Windhorst, ping Chen, Zou, Xia, Jiang, Ma, Xue, Zhu, Cheng, Byun, Chen, Deng,
		Fan, Fang, Kong, Li, Lin, Lu, hsin Sun, shun Tsay, Xu, Yan, Zhao, \&
		Zheng}]{Wu2002}
	Wu, H., Burstein, D., Deng, Z., {et~al.} 2002, \aj, 123, 1364
	
	\bibitem[{Yim {et~al.}(2014)Yim, Wong, Xue, Rand, Rosolowsky, van~der Hulst,
		Benjamin, \& Murphy}]{Yim2014}
	Yim, K., Wong, T., Xue, R., {et~al.} 2014, \aj, 148, 127
	
	\bibitem[{Zaritsky {et~al.}(2019)Zaritsky, Behroozi, Peeples, Tuttle, Werk, \&
		Zhang}]{Zaritsky2019}
	Zaritsky, D., Behroozi, P., Peeples, M.~S., {et~al.} 2019, \baas, 51, 127
	
	\bibitem[{Zibetti {et~al.}(2004)Zibetti, White, \& Brinkmann}]{Zibetti2004a}
	Zibetti, S., White, S. D.~M., \& Brinkmann, J. 2004, \mnras, 347, 556
	
	\bibitem[{Zschaechner {et~al.}(2012)Zschaechner, Rand, Heald, Gentile, \&
		J{\'o}zsa}]{Zschaechner2012}
	Zschaechner, L.~K., Rand, R.~J., Heald, G.~H., Gentile, G., \& J{\'o}zsa, G.
	2012, \apj, 760, 37
	
\end{thebibliography}
%

\bibliographystyle{aa} 


\begin{appendix} 

\section{Radial surface brightness profiles (RSBP)}  \label{appendix:RSBP}

\begin{figure*}
\includegraphics[width=\hsize]{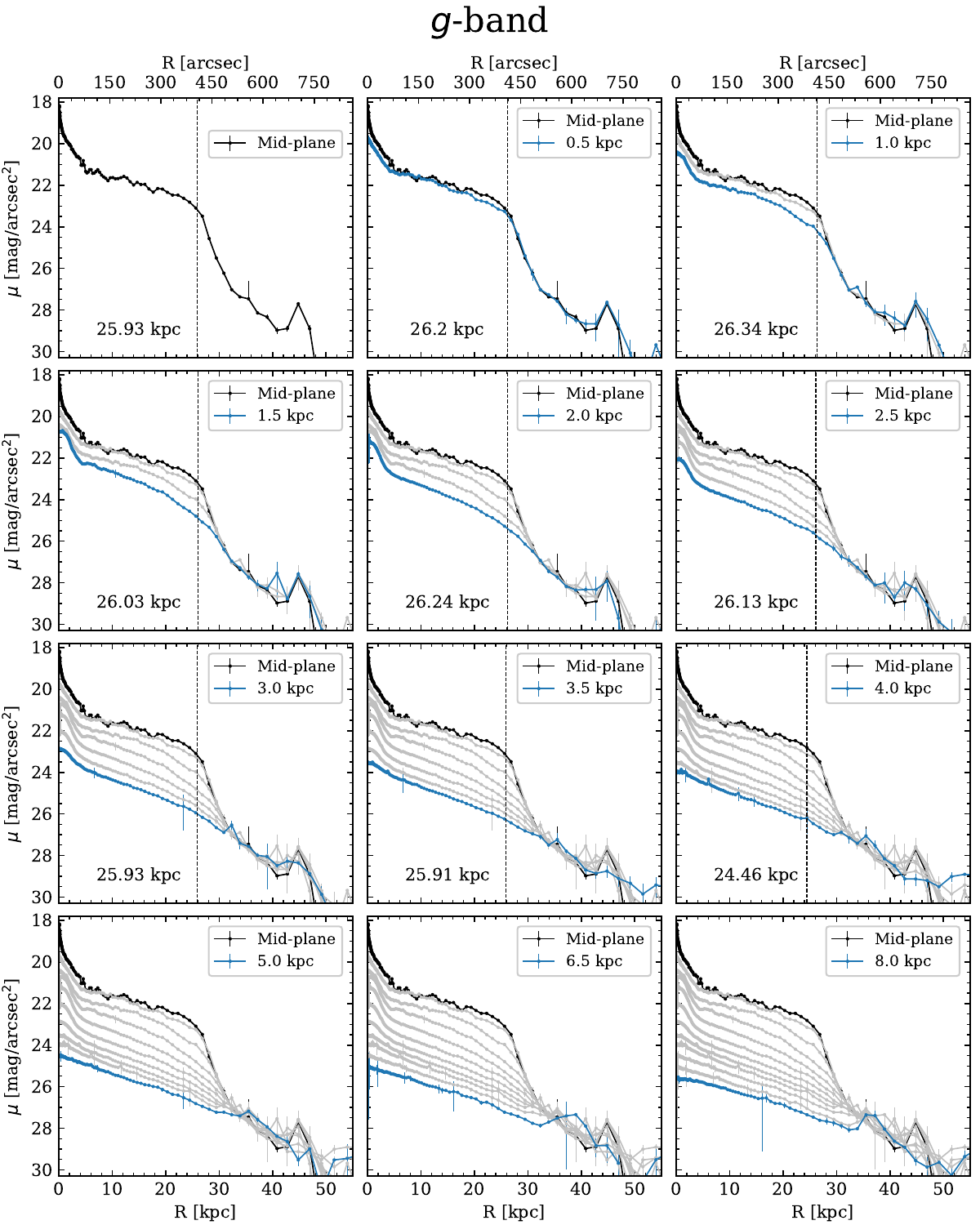}
\caption{Vertical development of the radial surface brightness profiles in the $g$ band for NGC~4565. The truncation radius is indicated for each panel. In all the cases the uncertainties are of the order of $2\%$ of the truncation radius value. The black line shows the mid-plane surface brightness profile. From top to bottom and from left to right, the panels show in blue the surface brightness profile at a given altitude above the galaxy mid-plane (indicated in the legend) and, in light grey, the surface brightness profiles plotted in the previous panels.}
\label{Fig:g-panel}
\end{figure*}

\begin{figure*}
\includegraphics[width=\hsize]{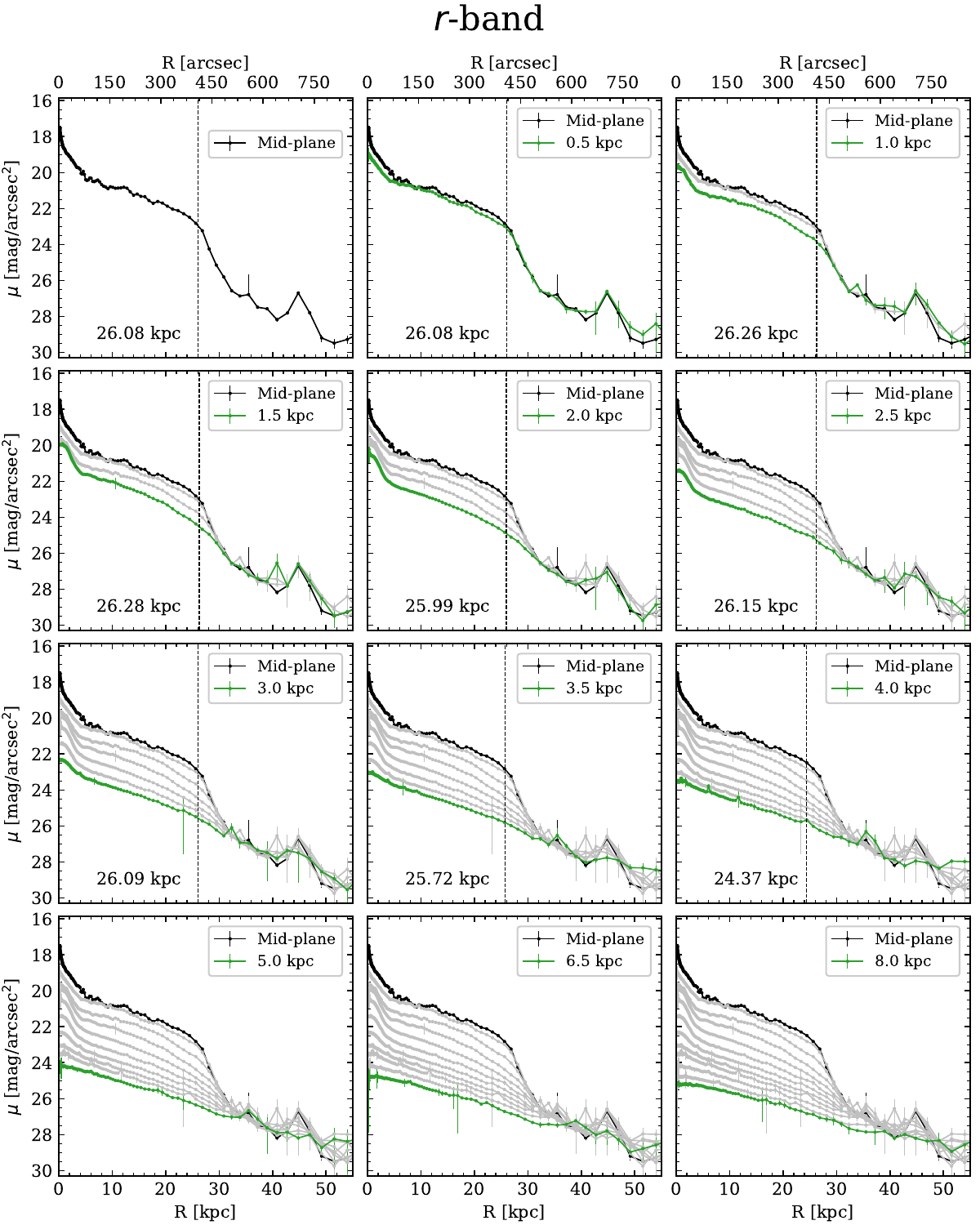}
\caption{Same than \ref{Fig:g-panel} but for $r$-band.}
\label{Fig:r-panel}
\end{figure*}

\begin{figure*}
\includegraphics[width=\hsize]{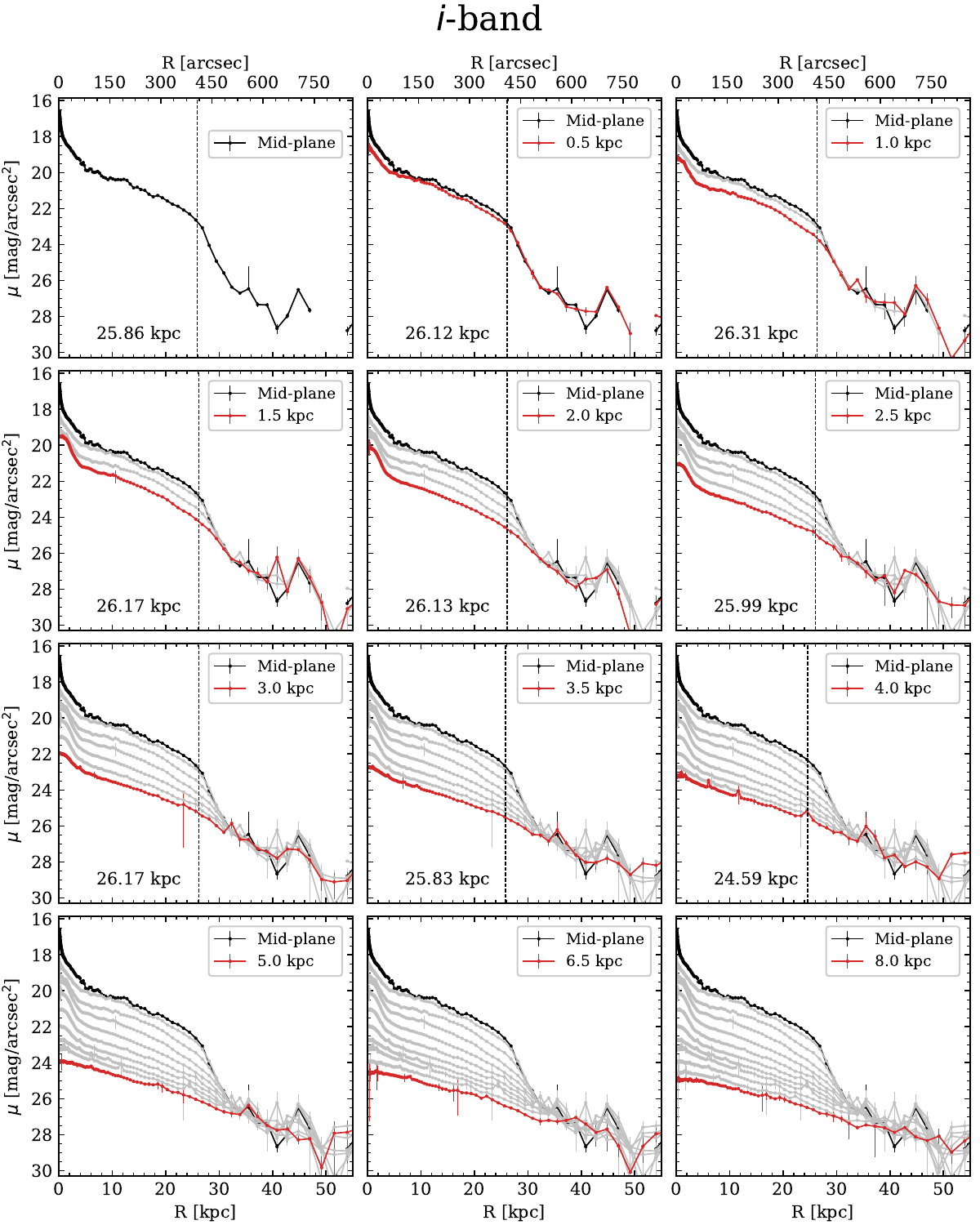}
\caption{Same than \ref{Fig:g-panel} but for $i$-band.}
\label{Fig:i-panel}
\end{figure*}

\begin{figure*}
\includegraphics[width=\textwidth]{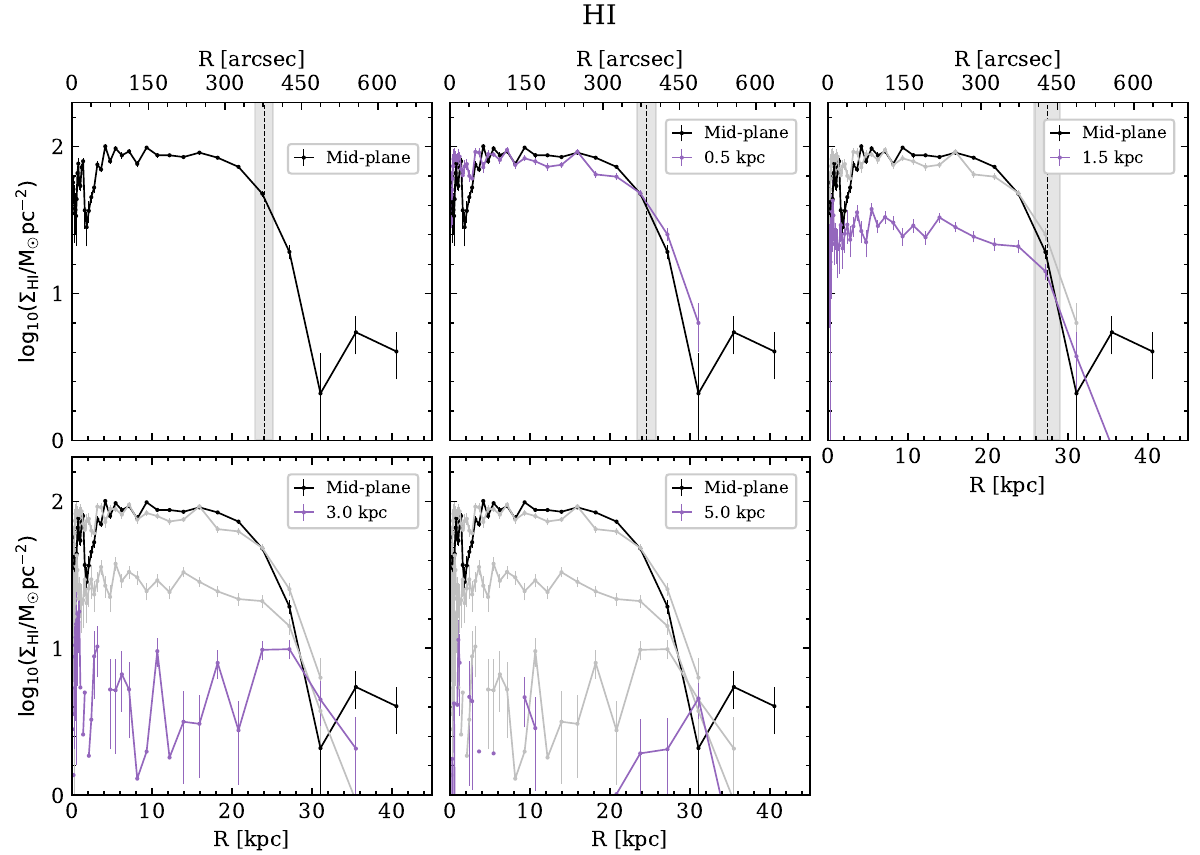}
\caption{Same than \ref{Fig:g-panel} but in this case it is shown the vertical development of the radial surface mass density profiles in M$_{\odot}$~pc$^{-2}$ extracted from the \ion{H}{i} integrated maps. The vertical line indicates the truncation radial position for each height above/below the galaxy mid-plane and the light grey shaded region is the associated uncertainty.}
\label{Fig:Hi-panel}
\end{figure*}

\begin{figure*}
\includegraphics[width=\hsize]{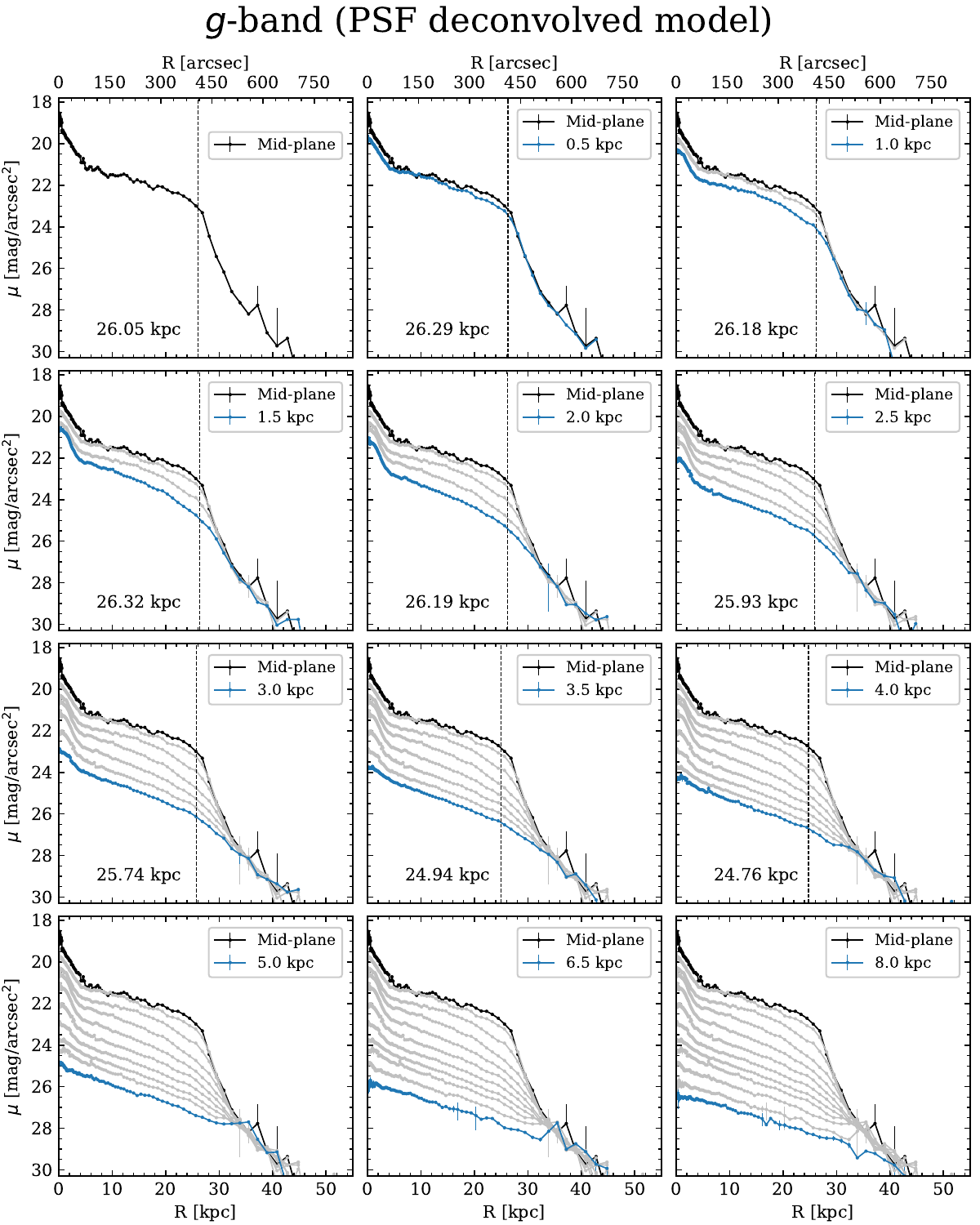}
\caption{Same than \ref{Fig:g-panel} but showing the vertical development of the radial surface brightness profiles extracted from the model of NGC~4565 in the $g$ band. The truncation radius is indicated for each panel. In all the cases the uncertainties are of the order of $2\%$ of the truncation radius value.}
\label{Fig:g-model-panel}
\end{figure*}

\begin{figure*}
\includegraphics[width=\hsize]{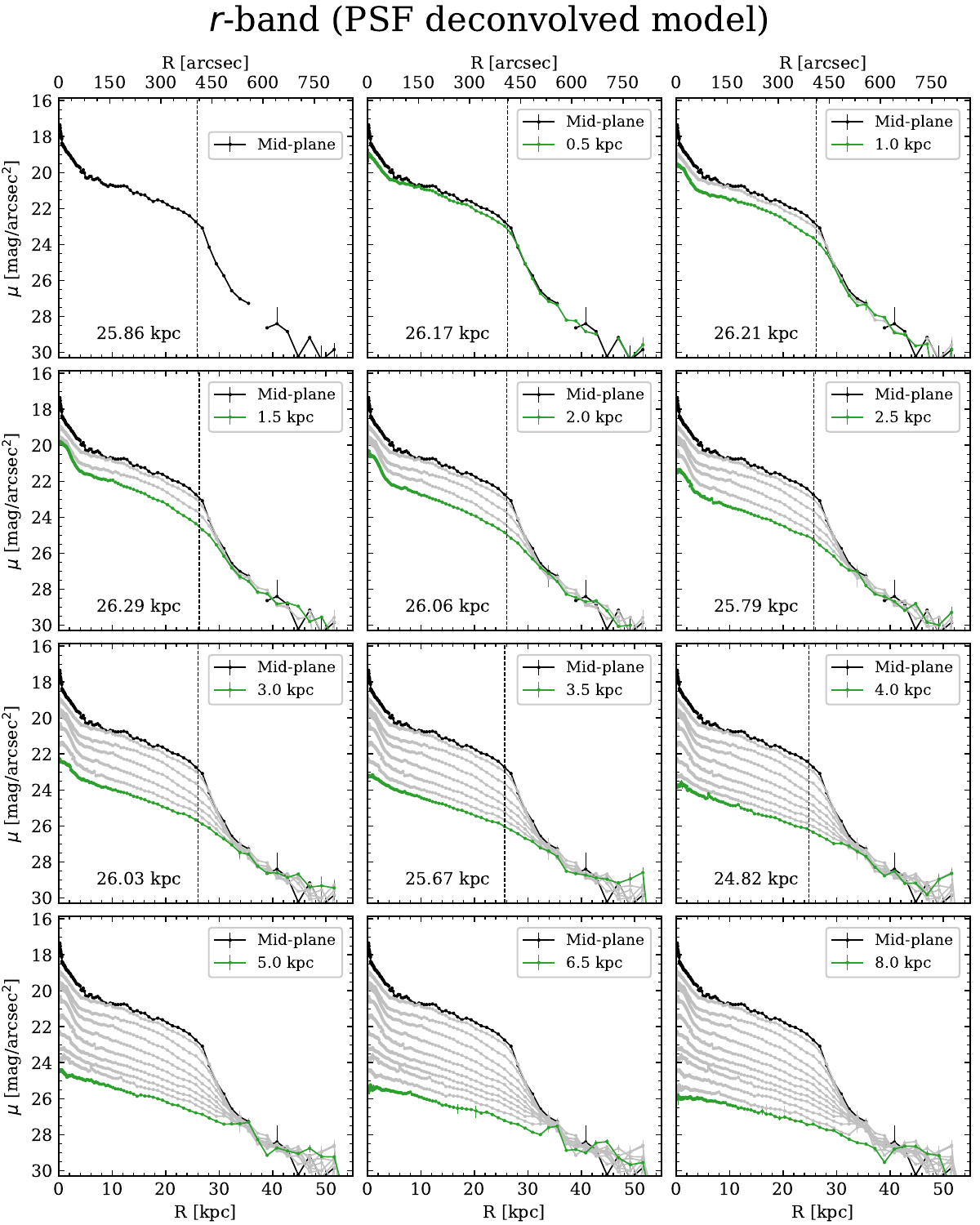}
\caption{Same than \ref{Fig:g-model-panel} but for $r$-band.}
\label{Fig:r-model-panel}
\end{figure*}

\begin{figure*}
\includegraphics[width=\hsize]{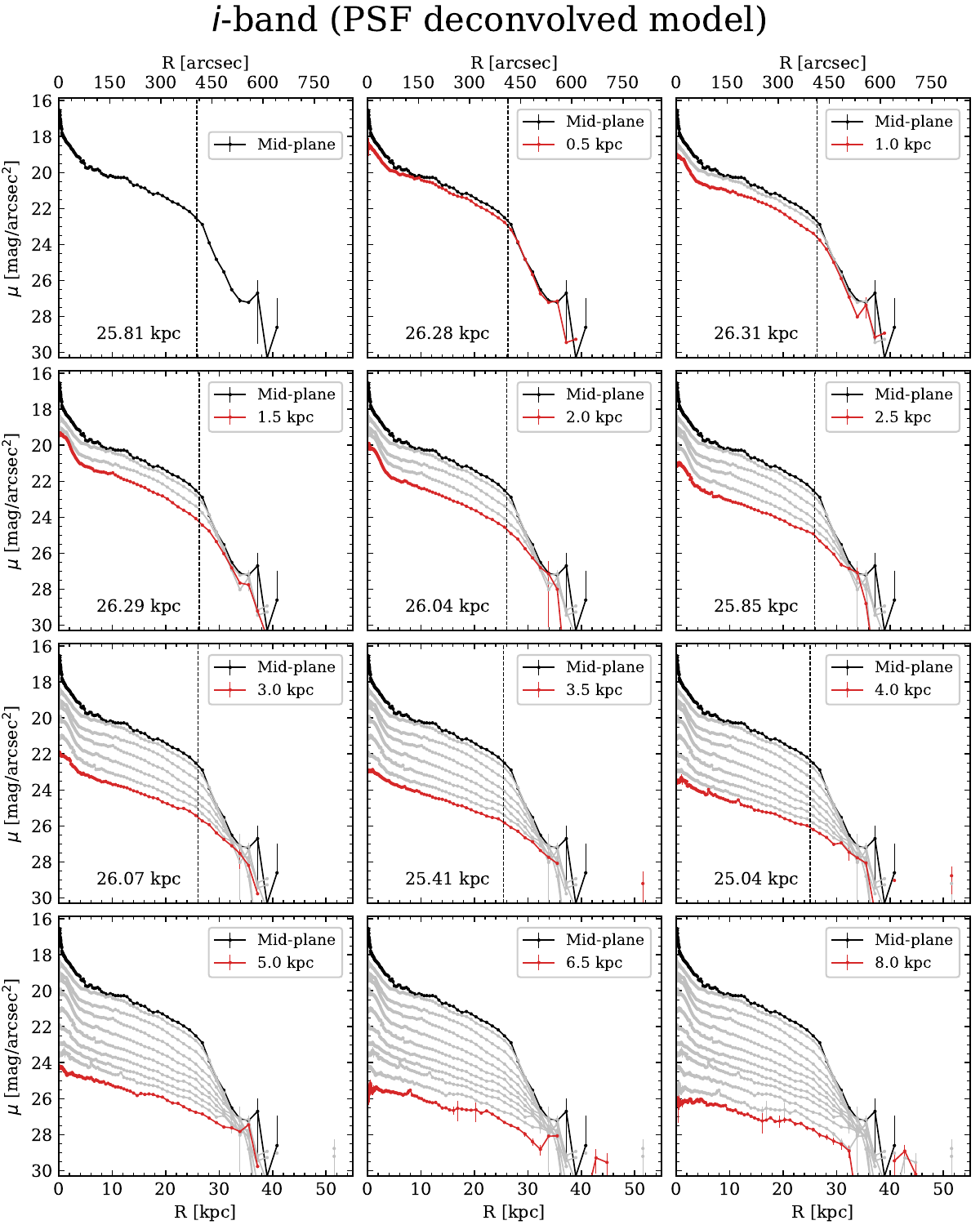}
\caption{Same than \ref{Fig:g-model-panel} but for $i$-band.}
\label{Fig:i-model-panel}
\end{figure*}

In Figs.~\ref{Fig:g-panel}, \ref{Fig:r-panel}, and \ref{Fig:i-panel}, we plot all the RSBPs of the observed data at different heights above/below the NGC~4565 mid-plane for the three ultra-deep $g$, $r$, and $i$ bands, respectively. Then, we show the same but for the \ion{H}{i} surface mass density profiles in Fig.~\ref{Fig:Hi-panel}. The RSBPs of the corresponding models in the $g$, $r$, and $i$ bands are shown in Figs.~\ref{Fig:g-model-panel}, \ref{Fig:r-model-panel}, and \ref{Fig:i-model-panel} respectively. 

Each figure shows the RSBP at all the different heights above/below the galaxy mid-plane in a given band. In this way, we explicitly illustrate the variations of the radial position of the truncation along the vertical axis as well as the profiles where the truncation is not detected. The black line always shows the mid-plane surface brightness profile. From top to bottom and from left to right, the panels show in colour the surface brightness profile at a given altitude above the galaxy mid-plane (indicated in the legend) and, in light grey, the surface brightness profiles plotted in the previous panels. When the truncation is detected by our algorithm, its radial position is also indicated with a value in kpc in the bottom left corner of each panel with a typical uncertainty of $2\%$ of the truncation radius value.

\section{NGC~4565 model}  \label{appendix:PSFmodels}

\begin{figure*}
\includegraphics[width=\textwidth]{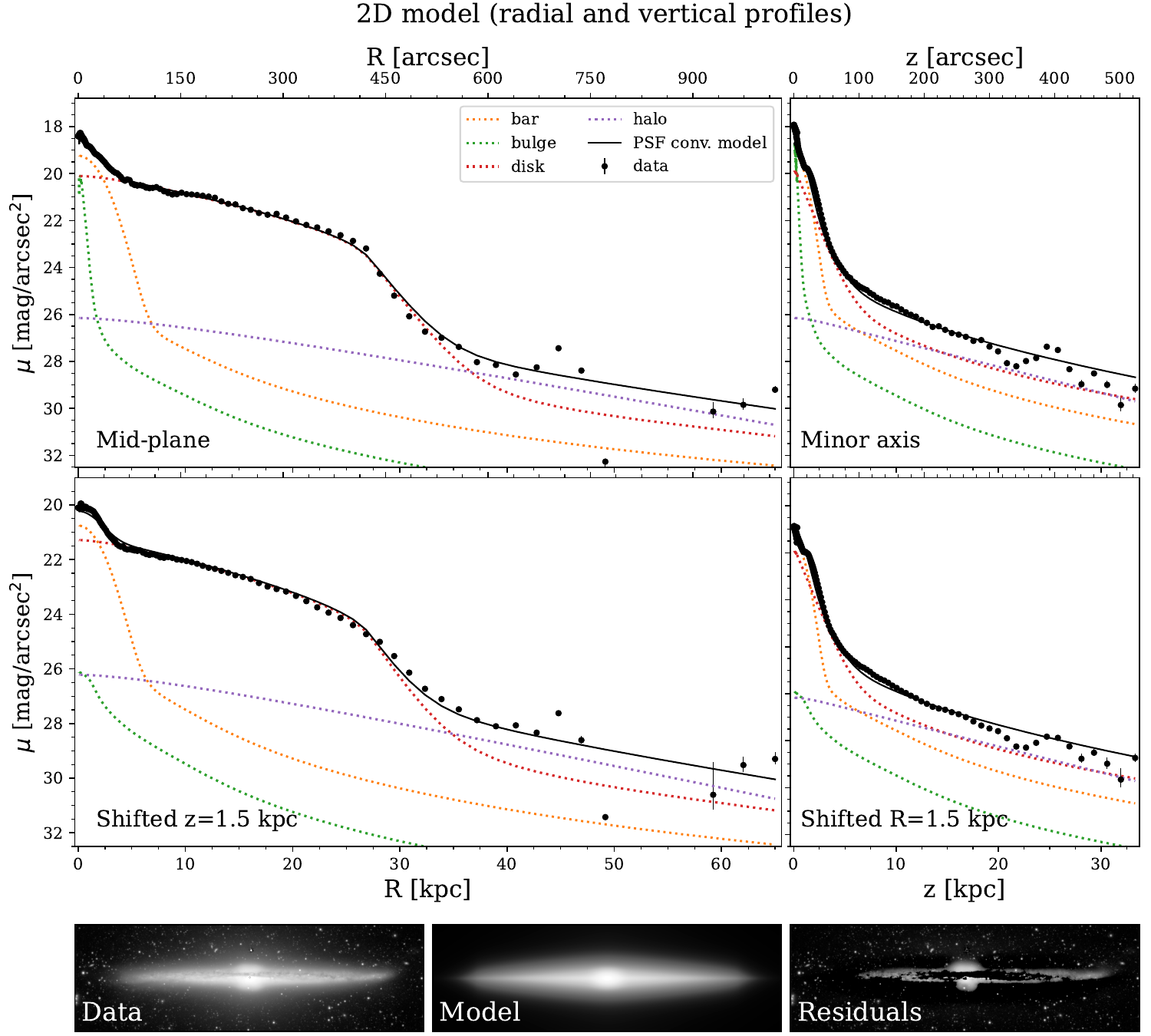}
\caption{\textbf{\textit{Top 4 panels:}} Radial (\textit{left}) and vertical (\textit{right}) surface brightness profiles of the observed data and the 2D galaxy model in $r$-band, convolved with the extended PSF (solid black line). We show the model galaxy components colour coded as in the legend, along with the observed data (black points). The two top panels show the central profiles in each case, while the other two panels are the profiles shifted 1.5~kpc from the ones above. \textbf{\textit{Bottom 3 panels:}} Images of the NGC~4565 observed data in $r$-band (\textit{left}), its model (\textit{centre}) obtained with \textsc{imfit} as explained in Sect.~\ref{subsection:NGC_PSFcorr}, and the associated residuals (\textit{right}). The model reproduces NGC~4565 and the residuals are minimal, just accounting for the dusty regions surrounding the mid-plane and some asymmetries.}
\label{Fig:model-profile}
\end{figure*}

Figure~\ref{Fig:model-profile} shows the details of the analytical 2D model of NGC~4565 and its components in comparison to the observed data in $r$-band. There is a good agreement between the observed data and the model in both, the 1D profiles and the 2D images. The results for the $g$ and $i$ band are similar. The full details on how to obtain this model are in Sect.~\ref{subsection:NGC_PSFcorr}.

\end{appendix}

\end{document}